\documentclass[journal]{IEEEtran}
\usepackage{amsmath,amsfonts,amsthm,amssymb,bm}
\usepackage{algorithmic}
\usepackage{array}
\usepackage[caption=false,font=normalsize,labelfont=sf,textfont=sf]{subfig}
\usepackage{textcomp}
\usepackage{stfloats}
\usepackage{url}
\usepackage{verbatim}
\usepackage{graphicx}
\usepackage{bbding}
\usepackage{multirow}
\usepackage{cite, bbm, xcolor}
\usepackage{refcount}
\usepackage{flafter}
\usepackage{makecell}
\usepackage{diagbox}
\usepackage{booktabs}
\usepackage{algorithm}

\newtheorem{theorem}{Theorem}
\newtheorem{proposition}{Proposition}
\newtheorem{lemma}{Lemma}
\newtheorem{corollary}{Corollary}

\theoremstyle{definition}

\usepackage{float}
\def\BibTeX{{\rm B\kern-.05em{\sc i\kern-.025em b}\kern-.08em
    T\kern-.1667em\lower.7ex\hbox{E}\kern-.125emX}}
\usepackage{balance}

\begin{document}
\title{Enabling Wireless Power Transfer (WPT) in Pinching Antenna Systems (PASS)}

\author{
        Deqiao Gan, Xiaoxia Xu, Xiaohu Ge, \textit{Senior Member, IEEE}, Yue Liu, and Yuanwei Liu, \textit{Fellow, IEEE}
        \thanks{D. Gan and X. Ge (Corresponding author) are with the School of Electronic Information and Communications, Huazhong University of Science and Technology, Wuhan 430074, Hubei, China. (e-mail: gandeqiao@hust.edu.cn, xhge@mail.hust.edu.cn).}
        \thanks{X. Xu is with the School of Electronic Engineering and Computer Science, Queen Mary University of London, London E1 4NS, U.K. (email: x.xiaoxia@qmul.ac.uk).}
        \thanks{Y. Liu is with the Department of Electrical and Electronic Engineering, The University of Hong Kong, Hong Kong (e-mail: yuanwei@hku.hk).}
        \thanks{Y. Liu is with the Faculty of Applied Sciences, Macao Polytechnic
        University, Macau, SAR, China (e-mail: yue.liu@mpu.edu.mo).}
}


\maketitle

\begin{abstract}
    A novel pinching antenna system (PASS) enabled wireless power transfer (WPT) framework is proposed, where energy harvesting receivers (EHRs) and information decoding receivers (IDRs) coexist. By activating pinching antennas (PAs) near both receivers and flexibly adjusting PAs' power radiation ratios, both energy harvesting efficiency and communication quality can be enhanced. A bi-level optimization problem is formulated to overcome the strong coupling between optimization variables. The upper level jointly optimizes transmit beamforming, PA positions, and feasible interval of power radiation ratios for power conversion efficiency (PCE) maximization under rate requirements, while the lower level refines power radiation ratio for the sum rate maximization. 
    Efficient solutions are developed for both two-user and multi-user scenarios. 
    1) For the two-user case, where an EHR and an IDR coexist, the alternating optimization (AO)-based and weighted minimum mean square error (WMMSE)-based algorithms are developed to achieve the stationary solutions of transmit beamforming, PA positions, and power radiation ratios. 2) For the multi-user case, a quadratic transform-Lagrangian dual transform (QT-LDT) algorithm is proposed to iteratively update PCE and sum rate by optimizing PA positions and power radiation ratios individually. Closed-form solutions are derived for both maximization problems. Numerical simulation results demonstrate that the proposed PASS-WPT framework significantly outperforms conventional MIMO and the baseline PASS with fixed power radiation, which demonstrates that: i) Compared to the conventional MIMO and baseline PASS, the proposed PASS-WPT framework achieves 81.45$\%$ and 43.19$\%$ improvements in PCE of EHRs, and ii) also increases the sum rate by 77.81$\%$ and 31.91$\%$ for IDRs.
\end{abstract}


\section{Introduction}
\IEEEPARstart{W}{ireless} power transfer (WPT) has emerged as a key enabling technology to support sustainable operation of energy-constrained wireless devices, particularly in next generation wireless networks \cite{pan2017,yuanwei2016wpt}. From an application perspective, WPT has been regarded as a fundamental technology to support large-scale Internet of things (IoT), wireless sensor networks (WSNs) \cite{pan2017,wang2016}, and unmanned aerial vehicle (UAV) networks \cite{2018uav}, where devices are energy-constrained but require long-term and stable operation. These promising applications highlight the crucial role of WPT in future next generation wireless communications \cite{tang2018wpt,2018uav}.
By delivering energy to low-power terminals, WPT can significantly prolong network lifetime and reduce reliance on frequent battery replacement \cite{Ioannis2014WPT}. However, conventional WPT architectures are fundamentally challenged by severe propagation loss and the difficulty of focusing power beaming precisely on distributed energy receivers. In high-frequency bands such as millimeter-wave (mmWave) and terahertz (THz), which are essential to support ultra-high data rates and device densities, radio waves experience severe propagation loss, high susceptibility to blockage, and strong spatial selectivity \cite{yang2025pinching,kit2021radio}.
Consequently, the efficiency of delivering stable energy to spatially distributed receivers is drastically reduced. These limitations motivate the exploration of novel physical-layer architectures that can reshape the propagation environment and provide flexible spatial control to achieve efficient wireless energy delivery \cite{SWIPTtutorial2015}.

Current WPT systems mostly depend on the massive multiple-input multiple-output (MIMO) system to increase the efficiency, especially in high-frequency systems \cite{heath2018foundations}. Massive MIMO arrays have been widely employed to compensate for path losses through highly directional beamforming \cite{2013MIMO}. However, the rigid antenna structures and static beam architectures in massive MIMO limit their ability to adapt to user mobility and dynamic propagation conditions. To enhance spatial adaptability, several flexible antenna solutions were explored, including reconfigurable intelligent surfaces (RISs) \cite{yuanwei2021RIS,tang2021RIS}, fluid antennas \cite{kit2021fluid,new2024fluid}, and movable antennas \cite{zhu2024movable,zhu2025movable}. Although these approaches improve propagation controllability, they still face practical challenges in scalability, integration, and real-time adaptability. Recently, the pinching antenna system (PASS) has emerged as a new form of flexible antenna technology with greater spatial scalability and reconfigurability \cite{2022NTTDOCOMO,ding2024pass}. PASS employs dielectric waveguides equipped with pinching antennas (PAs) that can be dynamically activated, deactivated, and repositioned along the waveguide \cite{yuanwei2025pass,gan2025NOMAPASS}. By enabling in-waveguide propagation and controllable pinching radiation, PASS not only reduces large-scale path loss and improves transmission efficiency \cite{liu2024path,chu2024propagation}, but also supports robust LoS connections and coverage adaptation to dynamic environments \cite{wang2025pass,xu2025pass2}. This unique spatial adaptability further leads to the paradigm of \emph{pinching beamforming}, where both large-scale attenuation and phase of the transmitted signal can be flexibly reconfigured \cite{ouyang2025pass,xu2025antenna}, offering advantages in adjustable positioning, real-time beam adaptation, and reliable blockage mitigation \cite{ding2025losblockage}. Such reconfigurability allows PASS to create effective LoS links and mitigate interference, thereby offering strong spatial controllability. Existing research has demonstrated that PASS significantly improves communication-oriented performance metrics such as sum rate and signal-to-interference-and-noise ratio (SINR) \cite{gan2025NOMAPASS}. Therefore, PASS exhibits these unique advantages:
\textit{1) Path-loss control and LoS support}: The adjustment of PAs allows PASS to dynamically control large-scale path loss while maintaining strong LoS links, significantly reducing the risk of high-frequency link blockage \cite{ding2025nomapass2}.
\textit{2) Adjustable position and beamforming}: PAs can be activated along the dielectric waveguide according to users' distributions and demands, enabling unprecedented adaptability over large physical areas. By activating PAs at the desired positions, PASS can achieve adjustable beamforming.

Building on these advantages, the potential of PASS supports this paradigm towards high-power WPT. The authors of \cite{passSWIPT} proposed the PASS framework aided by wireless information and power transfer (SWIPT) to maximize the sum rate. However, enabling efficient WPT in PASS has not been fully explored, since the integration of communication-oriented beamforming and power radiation mechanisms introduces several new research challenges:
\begin{itemize}
    \item \emph{Communication-WPT tradeoff}: Conventional PASS mainly targets communication quality. However, PASS-WPT needs to simultaneously enhance the power conversion efficiency (PCE) for energy harvesting receivers (EHRs) and improve the sum rate of information decoding receivers (IDRs). Moreover, the dual functionality introduces non-trivial trade-offs that maximum data rate may not align with the harvested energy \cite{george2022wpt}, i.e., power \cite{park2011wpt}. Designing a unified framework that balances PCE and communication quality under dynamic user demands is still a challenge.
    \item \emph{Adjustable power radiation control}: Previous studies typically assume equal or fixed power radiation across activated PAs, which oversimplifies practical scenarios \cite{xu2025pass2,wang2025pass,2025powerradiation}. The fundamental equal and proportional power models of PASS were constructed in \cite{xu2025pass2} by configuring the coupling length of each PA. Then, an adjustable power radiation model was investigated in \cite{wang2025pass} by flexibly tuning the waveguide-PA spacing.
    Moreover, the authors in \cite{2025powerradiation} optimized power radiation under PA motion costs.
    Different from these communication-centric designs, the power radiation adjustment in PASS-WPT is crucial to balance between EHRs' PCE and IDRs' data rate, which has rarely been investigated. 
    \item \emph{High-coupled optimization and interference management}: WPT must simultaneously mitigate inter-user interference in communication and guarantee sufficient energy harvesting \cite{tang2019pce,tang2018wpt}. This creates a highly coupled optimization problem, where PA position, transmit beamforming, and power allocation jointly affect the overall performance. Developing scalable optimization frameworks while ensuring both energy efficiency and fairness across users remains a fundamental challenge.
\end{itemize}

To address the above issues, we propose a PASS-WPT framework to jointly optimize transmit beamforming, PA positions, and power radiation ratio. To tackle the strong coupling among these optimization variables, a bi-level optimization problem is formulated. The upper-level problem is formulated by jointly optimizing the transmit beamforming, PA position, and feasible interval of power radiation. Then, the lower-level problem refines the power radiation ratio for sum rate maximization problem.
Further, this paper considers both the two-user and multi-user cases. For the two-user case, we design the optimization algorithms based on the alternating optimization (AO) and weighted minimum mean square error (WMMSE) to successively solve the above two maximization problems for the stationary solutions of transmit beamforming, PA positions and power radiation ratios. For the multi-user case, we propose a joint optimization algorithm based on quadratic transform-Lagrangian dual transform (QT-LDT) to maximize the above two problems by iterative PA position and power radiation ratio update steps, and subsequently derives the closed-form transmit beamforming, PA discrete positions and power radiation ratios. The key contributions are as follows:

\begin{enumerate}
    \item We propose a downlink PASS-WPT framework, where both EHRs and IDRs coexist. By configuring power radiation ratios of PAs, the PASS-WPT can improve the energy harvesting efficiency and communication performance. We formulate a bi-level optimization problem to achieve tractable variable decoupling. Specifically, the upper-level PCE maximization problem jointly optimizes transmit beamforming, PA positions, and feasible interval of power radiation ratio. Thereafter, the lower-level sum rate maximization problem refines power radiation ratio. To solve the bi-level optimization problem, this paper considers both two-user and multi-user cases.
    \item For the two-user case, where an EHR and an IDR, we design the AO-based algorithm to decompose the complex impact of PA positions on the channel coefficients, resulting in several inequality constraints. Under these inequality constraints, we derive the upper and lower bounds of power radiation ratio by maximizing the PCE. With the above optimized variables and feasible interval of power radiation ratio, we propose the WMMSE-based algorithm to maximize the lower-level sum rate by further refining this ratio.
    \item For the multi-user case, we develop a unified algorithm based on QT-LDT to iteratively update the upper-level PCE and the lower-level sum rate maximization problems. To eliminate the nonconvexity of fractional PCE and logarithmic sum rate, we reformulate these two problems into block-wise convex subproblems. The algorithm proceeds in two steps, where i) PA position update step optimizes the variables and feasible interval of power radiation ratios by solving the convex subproblems, and ii) power radiation ratio update step refines these ratios. This design achieves convergence to closed-form solutions. 
    \item Numerical results verify the effectiveness of the proposed PASS-WPT framework and the developed algorithms, which demonstrates that: i) The proposed PASS-WPT framework achieves 81.45$\%$ and 43.19$\%$ improvements in PCE of EHRs compared to conventional MIMO and conventional PASS. ii) The proposed framework also increases the sum rate of IDRs by 77.81$\%$ and 31.91$\%$.
\end{enumerate}

\section{System Model}
\begin{figure}[!t]
    \centering
    \includegraphics[width=3.6in]{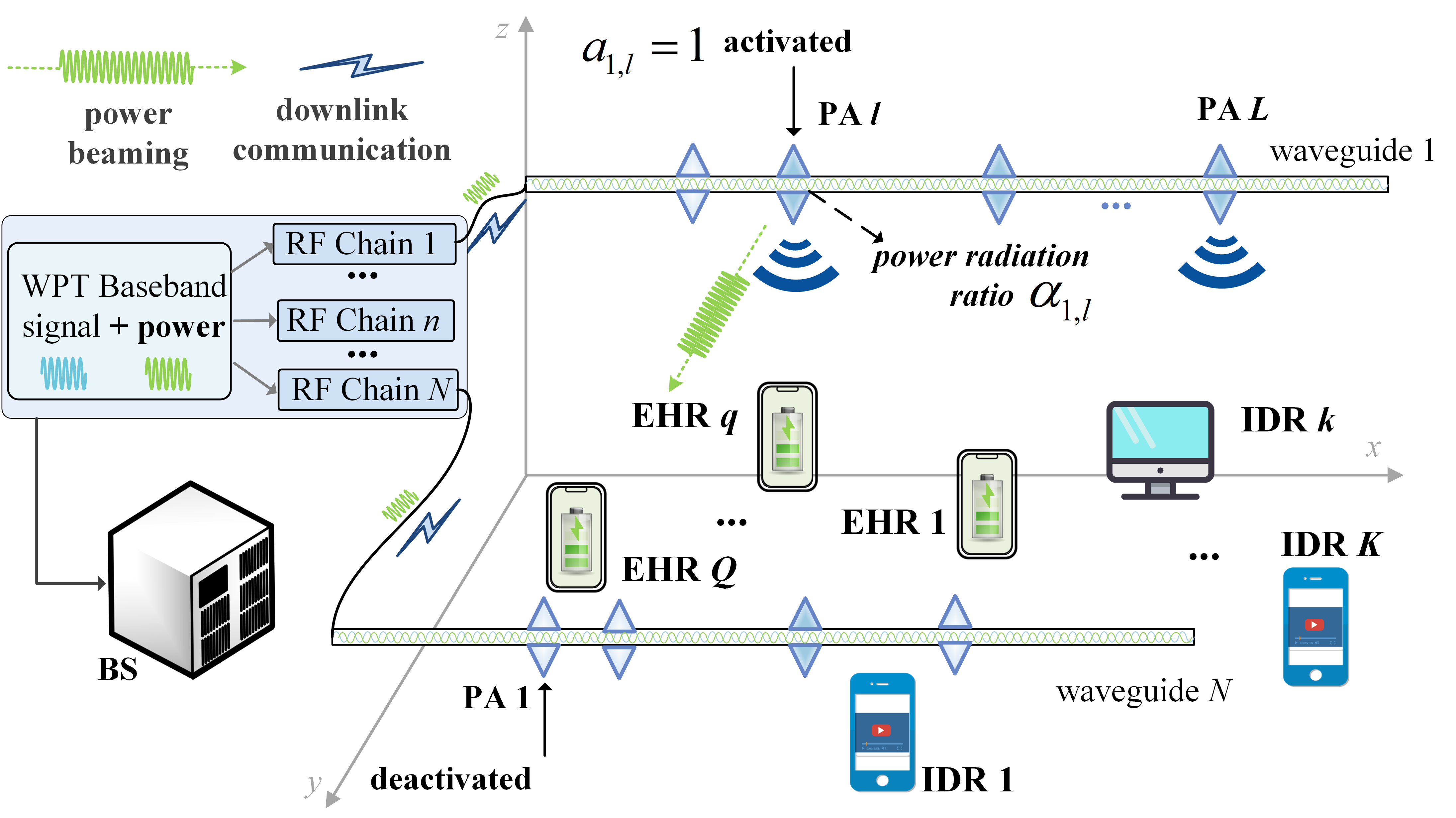}
    \caption{System model: The proposed PASS-WPT framework.}
    \label{systemmodel}
\end{figure}

We propose a PASS-WPT framework to maximize the power conversion efficiency by flexibly adjusting radiating powers of PAs, shown as Fig. \ref{systemmodel}.
We consider a downlink MIMO communication scenario, where a base station (BS) equipped with an edge server (ES) serves $K$ IDRs, namely communication users, and $Q$ EHRs, $k \in \mathcal{K}=[1,K]$, $q \in \mathcal{Q}=[1,Q]$. 
Specifically, PASS framework provide $N$ dielectric waveguides, pinched along flexible pinching antennas PAs $l \in \mathcal{L}=[1,L]$ of each waveguide $n \in \mathcal{N}={1,\dots,N}$. These PAs are activated by small particles, e.g., plastic pinches, to radiate signals from the waveguide into free space. Without loss of generality, these waveguides are deployed parallel to the $x$-axis at the height of $d_{\text{z}}^0$ with the feed point $\bm{\psi}^{\text{w}} = [0,(n-1)d_{\text{y}}^0,d_{\text{z}}^0]$, $d_{\text{y}}^0 \in [0,d_{\text{y}}^{\max}]$, which is the maximum distance of the waveguide along $y$-axis.
We denote the location of PA $l \in \mathcal{L}=\left\{1,...,L\right\}$ on waveguide $n$ as ${\bm{\psi}}^{\text{PA}}_{n,l}=[x_{n,l},(n-1)d_{\text{y}}^0,d_{\text{z}}^0]$, where $x_{n,l}$ is the location of PA along $x$-axis. Let $\mathbf{x}_{n}=[x_{n,1},x_{n,2},...,x_{n,L}]^T \in \mathbb{R}^{L \times 1}$ represent the x-axis locations of PAs on the waveguide $n$ with $0 \le x_{n,1} < x_{n,2} < x_{n,3} < ... < x_{n,L} \le x^{\max}$, $\forall n \in \mathcal{N}$, where $x^{\max}$ is the maximum length of rectangular service area along $x$-axis. And the position matrix of PAs is denoted by $\mathbf{X}=[\mathbf{x}_{1},\mathbf{x}_{2},...,\mathbf{x}_{N}] \in \mathbb{R}^{L\times N}$.
For a fixed discrete $\mathbf{X}$ and waveguide $n$ with feasible interval $[0,x^{\max}]$ and minimum spacing $\Delta$, pick grid step $\delta>0$ and define
\begin{equation}
    \begin{aligned}
        \mathcal{X} \! \triangleq \! \{x^{(n)}_j\!=b\Delta\!:\!b=0,1,\dots,j_n{-}1,\, n\in \mathcal{N}\},
        j_n \!\triangleq \! \Big\lfloor \tfrac{x^{\max}}{\Delta}\Big\rfloor \! +\! 1.
    \end{aligned}
    \label{discrete_x}
\end{equation}

We assume each IDR and EHR have the same height of 0 and same moving range along $x$-axis and $y$-axis, denoted by the three-dimensional Cartesian coordinate $\bm{\psi}^{\text{ID}}_k = [x^{\text{ID}}_k,d^{\text{ID}}_k,0]$, $d^{\text{ID}}_k \in [0,d_{\text{y}}^{\max}]$, $\bm{\psi}^{\text{EH}}_q = [x^{\text{EH}}_q,d^{\text{EH}}_q,0]$, $d^{\text{EH}}_q \in [0,d_{\text{y}}^{\max}]$.

\subsection{PASS Communication Model}

\subsubsection{Power Radiation Model}
We construct the adjustable power radiation model to flexibly control radiated powers by coupled mode theory \cite{coupledmodetheory1991}. Specifically, we control the coupling strength by configuring waveguide-antenna spacing. Since each power radiation ratio corresponds to a spacing value, the power radiation ratio can be optimized directly \cite{xu2025pass2}. Thus, the power radiation ratio $\alpha_{n,l}$ of PA $l$ along waveguide $n$ can be denoted by
\begin{equation}
    \label{powerradiatedratio}
    \alpha_{n,l} = a_{n,l}\sin\left(\epsilon_{l,n}d^{\max}_{\text{y}}\right)\prod_{i=1}^{l-1}\sqrt{1-a_{n,i}\sin^{2}\left(\epsilon_{l,n}d^{\max}_{\text{y}}\right)},
\end{equation}
where the power radiation ratio matrix is $\bm{\Lambda} = \text{diag}(\bm{\alpha}_1,\bm{\alpha}_2,\dots,\bm{\alpha}_N) \in \mathbb{R}^{NL \times N}, \forall n \in \mathcal{N}$, $\bm{\alpha}_n = [\alpha_{n,1},\alpha_{n,2},...,\alpha_{n,L}]^T \in \mathbb{R}^{L\times 1}$ and satisfies $\sum_{l\in \mathcal{L}}{\alpha_{n,l}}^2 \le 1$, $\forall l \in \mathcal{L}$. And $a_{n,l}=1$ means PA $l$ along waveguide $n$ is activated, and $a_{n,l}=0$ means deactivation.
The binary PA activation matrix $\text{diag}(\mathbf{a}_1,\mathbf{a}_2,\dots,\mathbf{a}_N) \in \mathbb{R}^{NL \times N}, \forall n \in \mathcal{N}$ comprises PA activation vector of each waveguide $\mathbf{a}_n=[a_{n,1},a_{n,2},\dots,a_{n,L}]^T \in \mathbb{R}^{L\times 1}$. The coupling coefficient $\epsilon_{l,n}$ measures the power exchange from waveguide $n$ to PA $l$ based on \cite{coupledmodetheory1991}. Thus, unlike the conventional works of fixed the radiation powers, we can adaptively control the radiated power of PASS by configuring power radiation ratio matrix. 
We assume lossless in-waveguide propagation since the in-waveguide attenuation can be ignored \cite{gan2025NOMAPASS,ouyang2025pass,ding2024pass,pass2025tutorial}. The baseband signals are multiplexed and passed through the RF chain and then fed into the waveguide as $\tilde{s}_k$, which satisfies $\mathbb{E}[\tilde{s}^H_k\tilde{s}_k]=1$.
The in-waveguide response matrix from the feed point to the corresponding PAs can be denoted by $\mathbf{G}(\mathbf{X}) = \text{diag}(\mathbf{g}_1, \mathbf{g}_2,\dots, \mathbf{g}_N) \in \mathbb{C}^{NL \times NL}$, consists of response vectors $\mathbf{g}_n = [g(x_{n,1}), g(x_{n,2}),\dots, g(x_{n,L})]^T \in \mathbb{C}^{L \times 1}$,
\begin{equation}
    \label{responsevector}
    g(x_{n,l})= \alpha_{n,l} e^{-i\frac{2\pi}{{\lambda}_g}\|{\bm{\psi}}^{\text{w}}-{\bm{\psi}}_{n,l}^{\text{PA}}\|}=\alpha_{n,l} e^{-i\frac{2\pi}{{\lambda}_g}x_{n,l}}, \forall l \in \mathcal{L},
\end{equation}
where $\lambda_g=\frac{\lambda}{n_{\text{eff}}}$ is the guided wavelength, $n_{{\text{eff}}}$ is the effective refractive index of the dielectric waveguide.

\subsubsection{Signal Model}
In this paper, we adopt a practical assumption that only the line-of-sight (LoS) components between each user and the BS are considered, while the non-LoS (NLoS) components are neglected due to their significantly weaker power compared to LoS paths.
Consequently, the channel between PA $l$ on the waveguide $n$ and IDR $k$ is modeled using a LoS-dominant framework, specifically employing a geometric free-space spherical propagation model to characterize the channel matrix ${\mathbf{h}}^H_{k}(\mathbf{X})=[{\mathbf{h}}^H_{1,k},{\mathbf{h}}^H_{2,k},\dots,{\mathbf{h}}^H_{N,k}] \in \mathbb{C}^{1\times NL}$, stacking all channel vectors ${\mathbf{h}}^H_{n,k}(\mathbf{x}_n) = [h_{n,l,k}^{H}(x_{n,1}),h_{n,l,k}^{H}(x_{n,2}),\dots,h_{n,l,k}^{H}(x_{n,L})] \in \mathbb{C}^{1 \times L}$, which comprises
\begin{equation}
    \label{channelelement_k}
    h_{n,l,k}^{H}(x_{n,l})=\frac{\eta e^{-i{\kappa}\|{\bm{\psi}}_{k}^{\text{ID}}-{\bm{\psi}}_{n,l}^{\text{PA}}\|}}{\|{\bm{\psi}}_{k}^{\text{ID}}-{\bm{\psi}}_{n,l}^{\text{PA}}\|},
\end{equation}
where $\kappa=\frac{2\pi}{\lambda}$ denotes the wave-domain number, $\lambda$ is the wavelength, the constant $\eta=\frac{c}{4\pi f_c}$ depends on the speed of light $c$ and carrier frequency $f_c$, and $\|{\bm{\psi}}_{k}^{\text{ID}}-{\bm{\psi}}_{n,l}^{\text{PA}}\| = \sqrt{(x_{n,l}-x^{\text{ID}}_k)^2 + [{(n-1)d_{\text{y}}^0-d^{\text{ID}}_k}]^2 + (d_{\text{z}}^0)^2}$.

After the in-waveguide propagation, the signal transmitted from the PAs along waveguide $n$ to IDR $k$ by transmit beamforming matrix $\mathbf{W} = [\mathbf{w}_1,\mathbf{w}_2,\dots,\mathbf{w}_K] \in \mathbb{C}^{N\times K}, \forall k \in \mathcal{K}$, which can be expressed as
\begin{equation}
    \label{transmitsignal_k}
    \mathbf{s}_{k} = \mathbf{G} \bm{\Lambda} \mathbf{w}_k \tilde{s}_k \in \mathbb{C}^{NL\times 1},
\end{equation}
where $\mathbf{w}_k = [w_{1,k},w_{2,k},\dots,w_{N,k}]^T \in \mathbb{C}^{N\times 1}$ represents the transmit beamforming vector for IDR $k$.

The received signal of IDR $k$ can be written as
\begin{equation}
    \label{receivedsignal_k}
    \begin{aligned}
        y_k & = \mathbf{h}^H_k \mathbf{s}_{k} + z_k \\
        & = \underbrace{\mathbf{h}^H_k\mathbf{G}\bm{\Lambda}\mathbf{w}_k \tilde{s}_k}_{\text{desired signal}} + \underbrace{\sum_{k^{'}\neq k}\mathbf{h}_{k}^H\mathbf{G}\bm{\Lambda}\mathbf{w}_{k^{\prime}}}_{\text{multi-user interference}}+z_k, \forall k \in \mathcal{K},
    \end{aligned} 
\end{equation}
where $z_k \sim \mathcal{CN}(0,\sigma^2)$ is the CSCG with zero mean, $\sigma^2$ is the noise power. The signal-to-interference-and-noise ratio (SINR) of IDR $k$ can be formulated as
\begin{equation}
    \label{SINR}
    \mathrm{SINR}_k=\frac{\left|\mathbf{h}_k^H \mathbf{G}\bm{\Lambda}\mathbf{w}_k\right|^2}{\sum_{k^{\prime}\neq k}\left|\mathbf{G}\bm{\Lambda}\mathbf{w}_{k^{\prime}}\right|^2+\sigma^2}.
\end{equation}

Then, the corresponding achievable rate is
\begin{equation}
    \label{rate_k}
    R_k=\log_2\left(1+\mathrm{SINR}_k(\mathbf{W},\mathbf{X})\right),
\end{equation}
and the system sum rate can be expressed as $\sum_{k \in \mathcal{K}}\log_2 \left(1+\mathrm{SINR}_k(\mathbf{W},\mathbf{X})\right)$.


\subsection{PASS-WPT Model}
We construct the WPT model to improve the energy efficiency of EHR to store the harvested energy. During the power beaming, the energy bearing signal, that is baseband signal $\tilde{s}_k, \forall k \in \mathcal{K}$, is multiplexed and transmitted to antennas of EHR $q$ via near-field physical channel, i.e., spherical-wave. The channel vector of EHR $q$ can be written as ${\mathbf{h'}}^H_{n,q}(\mathbf{x}_n) = [{h'}_{n,l,q}^{H}(x_{n,1}),{h'}_{n,l,q}^{H}(x_{n,2}),\dots,{h'}_{n,l,q}^{H}(x_{n,L})] \in \mathbb{C}^{1 \times L}$, which consists of
\begin{equation}
    \label{channelelement_q}
    {h'}_{n,l,q}^{H}(x_{n,l}) = \frac{\eta e^{-i{\kappa}\|{\bm{\psi}}_{q}^{\text{EH}}-{\bm{\psi}}_{n,l}^{\text{PA}}\|}}{\|{\bm{\psi}}_{q}^{\text{EH}}-{\bm{\psi}}_{n,l}^{\text{PA}}\|},
\end{equation}
and hence, the channel matrix between PAs and the EHR $q$ can be expressed as ${\mathbf{h}'}^H_{q}(\mathbf{X})=[{\mathbf{h}'}^H_{1,q},{\mathbf{h}'}^H_{2,q},\dots,{\mathbf{h}'}^H_{N,q}] \in \mathbb{C}^{1\times NL}$. The instantaneous signal transmitted by PAs along one waveguide is equal, thus, the received signal of EHR $q$ is given as
\begin{equation}
    \label{receivedsignal_q}
    y_q = {\mathbf{h}'}^H_q \mathbf{G} \bm{\Lambda} \mathbf{w}_k \tilde{s}_k + z_q, \forall k \in \mathcal{K},
\end{equation}
where $z_q \sim \mathcal{CN}(0,\sigma^2)$.
The EHR first captures the incident RF energy via its receiving antenna, then delivers the signal through an impedance matching network to a multi-stage rectifier circuit, where the high-frequency signal is converted into direct current (DC) power \cite{Ioannis2014WPT,george2022wpt}. After low-pass filtering and energy management, the harvested energy is stored in a battery or supplied to the load. We utilize the linear energy harvesting model to store the power \cite{tang2019pce}. The harvested power is proportional to the received RF power,
\begin{equation}
    \label{energy_q}
    P_q(\mathbf{X},\mathbf{W}) = \zeta_q \text{Tr}\left(\sum_{k=1}^{K} {\mathbf{h}'}^H_q \mathbf{G} \bm{\Lambda} \mathbf{w}_k{({\mathbf{h}'}^H_q \mathbf{G} \bm{\Lambda} \mathbf{w}_k)}^H\right),
\end{equation}
where $\zeta_q \in (0,1)$ is the efficiency of the power transducer to store the harvested power from RF to DC. The power conversion efficiency (PCE) is given as
\begin{equation}
    \label{powerefficiency}
    \text{PCE}(\mathbf{X},\mathbf{W}) = \frac{\sum_{q=1}^{Q} P_q(\mathbf{X},\mathbf{W})}{\sum_{k=1}^{K} \phi \|\mathbf{W}\|^2 + QP_{\text{C}}}, \forall q \in \mathcal{Q}, \forall k \in \mathcal{K},
\end{equation}
where $\phi$ is the reciprocal of the power amplifier drain efficiency, $\sum_{k=1}^{K} \|\mathbf{W}\|^2$ is the transmit power and $P_{\text{C}}$ is the circuit power required for supporting reliable power storage of each EHR.

\vspace{-0.2cm}
\subsection{Problem Formulations}
In the PASS-WPT framework, we aim to optimize the highly efficient power harvesting at EHRs, and meanwhile optimize the sum rate of all IDRs for next-generation wireless networks, where both high data rates and sustainable power supply are required. To mitigate the strong coupling among the spatial configuration, transmit beamforming, and power radiation, we adopt a bi-level problem design. The upper level problem maximizes PCE by optimizing transmit beamforming $\mathbf{W}$ and PA position $\mathbf{X}$ and deriving the feasible interval of power radiation ratio $\underline{\bm{\alpha}} \le \bm{\alpha} \le \bar{\bm{\alpha}}$. The lower level problem then refines $\bm{\alpha}$ within this interval to maximize the sum rate and yields closed-form solutions. For the upper level problem,
the PCE maximization problem can be formulated as
\begin{subequations}
    \label{PCE0}
    \begin{align}
        \mathbf{P}_0: \quad & \max_{\mathbf{W},\mathbf{X},\underline{\bm{\alpha}}, \overline{\bm{\alpha}}}\text{PCE}(\mathbf{W},\mathbf{X}),
            \label{minw_q}\\
            \mathrm{s.t.~} \quad & P_q(\mathbf{X},\mathbf{W},\bm{\alpha}) \ge P^{\min},k\in\mathcal{K},q\in\mathcal{Q},
            \label{powerharvest_low}\\
            & \mathrm{SINR}_{k}(\mathbf{W},\mathbf{X},\bm{\alpha}) \ge \gamma_{\min},k\in\mathcal{K},
            \label{SINRconstraint}\\
            & \sum_{k=1}^{K} \|\mathbf{W}\|^2 \le P^{\max},
            \label{transmitpower_high}\\
            & 0 \le \sum_{l\in \mathcal{L}}{\alpha_{n,l}}^2 \le 1, \forall l \in \mathcal{L} 
            \label{powerconstraints1}\\
            & x_{n,l}-x_{n,l-1} \ge \Delta, x_{n,l} \in \mathcal{X}, l\in\mathcal{L}, n\in\mathcal{N},
            \label{PA_x}
    \end{align}
\end{subequations}
where (\ref{powerharvest_low}) denotes the harvested power at each EHR must be higher than the minimum power threshold of storage $P^{\min}$, (\ref{SINRconstraint}) ensures the minimum SINR threshold $\gamma_{\min}$ of each IDR and derives the lower bound of $\bm{\alpha}$ for further refining it the of following lower-level problem, (\ref{transmitpower_high}) means the transmit power should not exceed the maximum transmit power budget $P^{\max}$, (\ref{powerconstraints1}) ensures the power radiation ratio variation range for all IDRs that it could not exceed the power of feed point on each waveguide deriving the upper-bound of $\bm{\alpha}$ for solving the following lower-level problem, constraint (\ref{PA_x}) denotes the minimum adjacent antennas space $\Delta$ to avoid mutual coupling and limits discrete PA positions on each waveguide.

For the lower level problem, we target to maximize the achievable sum rate for all IDRs by jointly power radiation ratios and PA positions, ensuring both high communication performance and sustainable energy supply. Thus, the sum rate maximization problem can be formulated as
\begin{subequations}
    \label{max_sumrate1}
    \begin{align}
            \mathbf{P}_1: \quad & \max_{\bm{\alpha}}\sum_{k \in \mathcal{K}}\log_2 \left(1+\mathrm{SINR}_k(\mathbf{W},\mathbf{X})\right),
            \label{maxsumratek}\\
            \mathrm{s.t.~} \quad & R_{k}(\mathbf{X}^{\star},\mathbf{W}^{\star},\bm{\alpha}) \ge R_{k}^{\mathrm{min}},k\in\mathcal{K},
            \label{rateconstraints}\\
            & \text{PCE}(\mathbf{X}^{\star},\mathbf{W}^{\star},\bm{\alpha}) \ge \varrho^{\min}, k\in\mathcal{K},q\in\mathcal{Q},
            \label{PCEconstraint}\\
            & \underline{\bm{\alpha}} \le \bm{\alpha} \le \bar{\bm{\alpha}},
            \label{alphaconstraint}
    \end{align}
\end{subequations}
where constraint (\ref{rateconstraints}) guarantees the minimum data rate of each IDR with the optimized transmit beamforming $\mathbf{W}^{\star}$ and PA position $\mathbf{X}^{\star}$, (\ref{PCEconstraint}) means each EHR should reach the minimum PCE threshold $\varrho$ as the IDRs are set by optimized $\mathbf{X}^{\star}$ and $\mathbf{W}^{\star}$, (\ref{alphaconstraint}) denotes the feasible interval of $\bm{\alpha}$ for finetuning.

The PCE maximization problem is highly non-convex and NP-hard due to the fractional and non-convex nature of the objective function, as well as strong coupling $\mathbf{G}\bm{\Lambda}\mathbf{w}$ between the transmit beamforming, PA position and bounds of power radiation ratio. Moreover, the sum rate maximization problem formulated above is also highly non-convex, as well as the same strong coupling from $\mathbf{G}\bm{\Lambda}\mathbf{w}$, which render the feasible set of $\bm{\alpha}$ non-convex. To efficiently solve these problems, we resort to an AO and augmented Lagrangian (AL)-based approach, iteratively updating the transmit beamforming, PA positions and power radiation ratios.


\section{Optimization Of PASS-WPT For Two-User Case}

In this section, we propose the joint beamforming and power radiation optimization algorithm to maximize the PCE and sum rate for two-user case, i.e., one EHR and one IDR. Furthermore, we derive the optimized solutions of the PCE maximization problem and the sum rate maximization problem, which are the transmit beamforming, PA positions and power radiation ratios.

\subsection{Upper-level Problem Solution of PCE Maximization}
\begin{algorithm}[!t]
    \caption{AO-based PCE Maximization Algorithm in PASS-WPT for Two-User Case}
    \label{alg:two_pce}
    \small
    \begin{algorithmic}[1]
    \REQUIRE $Q$, $K$, $N$, $L$, $\eta$, $\sigma^2$, $T$, tolerance $\epsilon_0$
    \ENSURE $\mathbf{w}^*,\,\mathbf{X}^*$
    \STATE \textbf{Initialize:} $\mathbf{X}^{(0)}$, $\bm{\alpha}^{(0)}$, $\mathbf{w}^{(0)}$ feasible with $\|\mathbf{w}^{(0)}\|^2\!\le\!p_0$ and $|\mathbf{h}(\mathbf{X}^{(0)})^H\mathbf{w}^{(0)}|^2\!\ge\!\gamma_{\min}\sigma^2$; $\beta_0^{(0)}\!\leftarrow\!0$; $i\!\leftarrow\!0$
    \STATE \textbf{Outer-Loop: (3-12)}
    \REPEAT
        \STATE \textbf{Transmit Beamforming Optimization:} (given $\mathbf{X}^{(i)}$, $\bm{\alpha}^{(i)}$, $\beta_0^{(i)}$): Dinkelbach
        (\ref{PCE00})
        \STATE \textbf{Inner-Loop: (6-9)}
        \STATE Transform the (\ref{PCE00}) into (\ref{PCE00_object})
        \STATE One-Dimensional Search: Search $\lambda_0$ to maximize objective under constraints from (\ref{transmit_1})-(\ref{maximum_p})
        \STATE Update (\ref{beta_0i})
        \STATE \textbf{PA Position Optimization:} with $\mathbf{w}^{(i+1)}$, $\bm{\alpha}^{(i)}$, $\beta_0^{(i)}$, update $\mathbf{X}^{(i+1)}$ in (\ref{optimal_x1}), $\beta^{(i+1)}$ in (\ref{beta_0i})
        \STATE $i \leftarrow i+1$
    \UNTIL{$\Big|\zeta_1|{\mathbf{h}'^H}\mathbf{G}\bm{\Lambda}\mathbf{w}^{(i)}|^2 - \beta_0^{(i)}\big(\varphi\|\mathbf{w}^{(i)}\|^2{+}P_{\text{C}}\big)\Big|<\epsilon_0$}
    \STATE \textbf{RETURN} $\mathbf{w}_1^*=\mathbf{w}^{(i)}$, $\mathbf{X}^*_{\text{two}}=\mathbf{X}^{(i)}$, \eqref{feasibleinterval_two}
\end{algorithmic}
\end{algorithm}

Since the PCE maximization problem is non-convex and fractional, the Dinkelbach approach aims to solve concave-convex fractional problems by continually shrinking the domain of the objective function to obtain an optimized solution. We utilize the Dinkelbach approach to transform (\ref{PCE0}) into the equivalence difference formula,
\begin{subequations}
    \label{PCE00}
    \begin{align}
        \nonumber
        \mathbf{P}_{0.0}: \ & \max_{\mathbf{X}, \mathbf{w}_1, \underline{\bm{\alpha}_1}, \overline{\bm{\alpha}_1}}{\sum_{q=1}^{Q} \! P_q(\mathbf{X},\mathbf{w}_1)} - \!\beta_0 \! \left({\sum_{k=1}^{K} \phi \|\mathbf{w}_1\|^2 \!+\! QP_{\text{C}}}\right),\\
        & = \max_{\mathbf{X}, \mathbf{w}_1}{ P_q(\mathbf{X},\mathbf{w}_1)} - \beta_0 \left({ \phi \|\mathbf{w}_1\|^2 + P_{\text{C}}}\right),
            \label{maxw_1}\\
            \mathrm{s.t.~} \ & \text{(\ref{powerharvest_low})-(\ref{PA_x})}, \quad k=K=1, q=Q=1,
    \end{align}
\end{subequations}
where $\beta_0$ is the auxiliary parameter for iterations and \eqref{SINRconstraint} is transformed to ensure the required minimum SNR threshold in two-user case. 
Furthermore, we adopt AO method \cite{2011AO} to solve the (\ref{PCE00}). For the two-user PASS-WPT, the PCE maximization can be globally solved via the Dinkelbach method combined with AO \cite{dinkelbach1967}.

During the iteration process, we assume the optimized variables $\mathbf{X}$, $\mathbf{w}_1$ is obtained at the $i$-th iteration. The auxiliary parameter $\beta_0^{(i+1)}$ can be expressed as
\begin{equation}
    \label{beta_0i}
    \begin{aligned}
        \beta_0^{(i+1)} & = \frac{ P_q(\mathbf{X}^{(i)},\mathbf{w}_1^{(i)})}{ \phi {(\|\mathbf{w}_1\|^2)}^{(i)} + P_{\text{C}}}\\
        & = \frac{\zeta_1 \text{Tr}\left( {\mathbf{h}'}^H_1 \mathbf{G} \bm{\Lambda} \mathbf{w}_1^{(i)}{({\mathbf{h}'}^H_1 \mathbf{G} \bm{\Lambda} \mathbf{w}_1^{(i)})}^H\right)}{ \phi {(\|\mathbf{w}_1\|^2)}^{(i)} + P_{\text{C}}},
    \end{aligned}
\end{equation}
where $\mathbf{X}^{(i)},\mathbf{w}^{(i)}$ denotes the optimized solutions at $i$-th iteration. In each iteration, given a fixed PA positions, the optimized transmit beamforming vector $\mathbf{w}_1^*$ is found via analytical one-dimensional search, and then the optimized PA position $\mathbf{X}^{*}$ is determined by maximizing the objective over all feasible $\mathbf{X}^{*}$. This iterative procedure converges to the global optimum. For this two-user case, since the optimized $\mathbf{w}_1^*$ must lie in the two-dimensional subspace spanned by $\mathbf{h}_1^H$ and ${\mathbf{h}'}^H_1$, we can formulate
\begin{equation}
    \label{transmit_1}
    \mathbf{w}_1 = \sqrt{p_0}\frac{\mathbf{h}_1^H+\lambda_0 {\mathbf{h}'}^H_1}{\|\mathbf{h}_1^H+\lambda_0 {\mathbf{h}'}^H_1\|},
\end{equation}
where $\lambda_0$ is a real scalar to be optimized and $p_0 = \|\mathbf{w}\|^2$ is the transmit power for this case.
Hence, the objective of (\ref{PCE00}) can be transformed into
\begin{equation}
    \label{PCE00_object}
    \begin{aligned}
        & { P_q(\mathbf{X},\mathbf{w}_1)} - \beta_0 \left({ \phi \|\mathbf{w}_1\|^2 + P_{\text{C}}}\right) \\
        & = p_0 \left[\frac{\left|{\mathbf{h}_1^H+\lambda_0 {\mathbf{h}'}^H_1}\right|^2}{\|\mathbf{h}_1^H+\lambda_1 {\mathbf{h}'}^H_1\|^2} - \beta_0 \left({ \phi + \frac{P_{\text{C}}}{p_0}}\right)\right],\\
        & \|\mathbf{w}_1\|^2 = p_0 \le P^{\max}, \quad p_0 \frac{\left|{\mathbf{h}_1^H+\lambda_0 {\mathbf{h}'}^H_1}\right|^2}{\|\mathbf{h}_1^H+\lambda_1 {\mathbf{h}'}^H_1\|^2} \ge \gamma_{\min}{\sigma}^2,
    \end{aligned}
\end{equation}
where $\gamma_{\min}$ is the minimum threshold SNR. For each $\lambda_0$ during the iterations, $p_0$ satisfies $P^{\max} \ge p_0 \ge \frac{\|\mathbf{h}_1^H+\lambda_1 {\mathbf{h}'}^H_1\|^2}{\left|{\mathbf{h}_1^H+\lambda_0 {\mathbf{h}'}^H_1}\right|^2}\gamma_{\min}{\sigma}^2$ and maximal feasible $p_0$ can be formulated as
\begin{equation}
    \label{maximum_p}
    p_0^*(\lambda_0) = \min \left\{P^{\max}, \frac{\|\mathbf{h}_1^H+\lambda_0 {\mathbf{h}'}^H_1\|^2}{\left|{\mathbf{h}_1^H+\lambda_0 {\mathbf{h}'}^H_1}\right|^2}\gamma_{\min}{\sigma}^2 \right\}^{-1}.
\end{equation}


Substitute this $p_0^*(\lambda_0)$ into the objective and perform a one-dimensional search over all feasible $\lambda_0$ to find the value $\lambda_0^*$ that maximizes the objective. The corresponding $\mathbf{w}_1^*$ can be expressed as
\begin{equation}
    \label{optimal_w1}
    \mathbf{w}^*_1 = \sqrt{p^*_0(\lambda^*_0, \mathbf{X}^*)}\frac{\mathbf{h}_1^H(\mathbf{X}^*) +\lambda^*_0 {\mathbf{h}'}^H_1(\mathbf{X}^*)}{\|\mathbf{h}_1^H+\lambda^*_0 {\mathbf{h}'}^H_1(\mathbf{X}^*)\|},
\end{equation}
and the optimized PA positions matrix can be given by
\begin{equation}
    \label{optimal_x1}
    \begin{aligned}
        \mathbf{X}^*_{\text{two}} = & \arg \max_{\mathbf{X}} \zeta_1 \text{Tr}\left( {\mathbf{h}'}^H_1 \mathbf{G} \bm{\Lambda} \mathbf{w}_1^{*}{({\mathbf{h}'}^H_1 \mathbf{G} \bm{\Lambda} \mathbf{w}_1^{*})}^H\right) \\
        & - \beta_0 \left({ \phi \|\mathbf{w}^*_1\|^2 + P_{\text{C}}}\right).
    \end{aligned}
\end{equation}

If $\alpha^2<AB$, then for $\lambda\!>\!0$ the IDR direction function $f(\lambda)\!=\!\frac{(A+\lambda\alpha)^2}{A+2\lambda\alpha+\lambda^2B}$ is strictly decreasing and the EHR function $g(\lambda)\!=\!\frac{(\alpha+\lambda B)^2}{A+2\lambda\alpha+\lambda^2B}$ is strictly increasing.

Define the auxiliary vectors for simplicity,
\begin{subequations}
    \label{c_d}
    \begin{align}
        \mathbf{c}_1\triangleq \big(\mathbf{h}_1(\mathbf{X})^{H}\mathbf{G}(\mathbf{X})\bm{\Lambda}(\bm{\alpha})\big)^{H},\\
        \mathbf{d}_1\triangleq \big(\mathbf{h}'_1(\mathbf{X})^{H}\mathbf{G}(\mathbf{X})\bm{\Lambda}(\bm{\alpha})\big)^{H}.
    \end{align}
\end{subequations}
        
\begin{corollary}[Monotonicity of direction functions of IDR and EHR]\label{lem:mono}
    Let $C=\|\mathbf{c}_1\|_2^2>0$, $D=\|\mathbf{d}_1\|_2^2>0$, and $\delta_0 = \Re\{\mathbf{c}_1^H\mathbf{d}_1\}\ge 0$.
    Define, for $\lambda_0 \ge 0$, direction functions of IDR and EHR as 
    \begin{equation}
        \label{f_ID0}
        f_{\text{ID}0}(\lambda_0)\triangleq\frac{(C+\lambda_0\delta_0)^2}{C+2\lambda_0\delta_0+\lambda_0^2 D},
    \end{equation}
    \begin{equation}
        \label{f_EH0}
        f_{\text{EH}0}(\lambda_0)\triangleq\frac{(\delta_0+\lambda_0 D)^2}{C+2\lambda_0\delta_0+\lambda_0^2 D}.
    \end{equation}

    If $\delta_0^2<CD$, $\mathbf{c}_1$ and $\mathbf{d}_1$ are not positively colinear, then $f_{\text{ID}0}(\lambda_0)$ is strictly decreasing on $(0,\infty)$ and $f_{\text{EH}0}(\lambda_0)$ is strictly increasing on $[0,\infty)$. If $\delta_0^2=CD$ positively colinear, then $f_{\text{ID}0}(\lambda_0)\equiv C$, $f_{\text{EH}0}(\lambda_0)\equiv D$ are constant for all $\lambda_0 \ge 0$.
\end{corollary}
    
    \begin{IEEEproof}
    Denote $q(\lambda_0)\triangleq C+2\lambda_0\delta_0+\lambda_0^2D=\|\mathbf{c}_1+\lambda_0\mathbf{d}_1\|_2^2>0$ for all $\lambda_0\ge 0$. 
    
    \emph{(i) Derivative of $f_{\text{ID}0}$.} Using the quotient rule, it can be expressed as
    \begin{equation}
        \label{derivative_IDR}
            f_{\text{ID}0}'(\lambda_0) = \frac{2(C+\lambda_0\delta_0)\,\lambda_0\,(\delta_0^2-CD)}{(C+2\lambda_0\delta_0+\lambda_0^2D)^2}.
    \end{equation}
    
    For $\lambda_0>0$, $C+\lambda_0\delta_0\ge C>0$ and $(C+2\lambda_0\delta_0+\lambda_0^2D)^2>0$, so the sign of $f_{\text{ID}0}'(\lambda_0)$ equals the sign of $(\delta_0^2-CD)$. Under $\delta_0^2<CD$, we have $f_{\text{ID}0}'(\lambda_0)<0$ for all $\lambda_0>0$, i.e., $f$ is strictly decreasing on $(0,\infty)$. (At $\lambda_0=0$, $f_{\text{ID}0}'(0)=0$ by the factor $\lambda_0$.)
    
    \emph{(ii) Derivative of $f_{\text{EH}0}$.} Similarly, we can get
    \begin{equation}
        \label{derivative_EHR}
        f_{\text{EH}0}'(\lambda_0)
        =\frac{2(\delta_0+\lambda_0 D)\,(CD-\delta_0^2)}{(C+2\lambda_0\delta_0+\lambda_0^2D)^2}.
    \end{equation}
    Since $\delta_0\ge 0$, $D>0$, and $\lambda_0\ge 0$, we have $\delta_0+\lambda_0 D\ge 0$; with $\delta_0^2<CD$ we obtain $CD-\delta_0^2>0$. Because $q(\lambda_0)^2>0$, it follows that $f_{\text{EH}0}'(\lambda_0)>0$ for all $\lambda_0\ge 0$, so $g$ is strictly increasing on $[0,\infty)$.
    
    \emph{(iii) Colinear boundary case.} If $\delta_0^2=CD$ and $\delta_0\ge 0$, it can be derived, i.e.,
    \begin{equation}
        \label{norm_cd}
        \|\mathbf{c}_1+\lambda_0\mathbf{d}_1\|_2^2
        =\big(\|\mathbf{c}_1\|_2+\lambda_0\|\mathbf{d}_1\|_2\big)^2,
    \end{equation}
    and
    \begin{subequations}
        \begin{align}
            (C+\lambda_0\delta_0)=\|\mathbf{c}_1\|_2\big(\|\mathbf{c}_1\|_2+\lambda_0\|\mathbf{d}_1\|_2\big),
            \label{C}\\
            (\delta_0+\lambda_0 D)=\|\mathbf{d}_1\|_2\big(\|\mathbf{c}_1\|_2+\lambda_0\|\mathbf{d}_1\|_2\big).
            \label{D}
        \end{align}
    \end{subequations}

    Therefore, we can obtain
    \begin{equation}
        \label{2f_ID0}
        f_{\text{ID}0}(\lambda_0)=\frac{\|\mathbf{c}_1\|_2^2(\|\mathbf{c}_1\|_2+\lambda_0\|\mathbf{d}_1\|_2)^2}{(\|\mathbf{c}_1\|_2+\lambda_0\|\mathbf{d}_1\|_2)^2}=C,
    \end{equation}
    \begin{equation}
        \label{2f_EH0}
        f_{\text{EH}0}(\lambda_0)=\frac{\|\mathbf{d}_1\|_2^2(\|\mathbf{c}_1\|_2+\lambda_0\|\mathbf{d}_1\|_2)^2}{(\|\mathbf{c}_1\|_2+\lambda_0\|\mathbf{d}_1\|_2)^2}=D,
    \end{equation}
    which both are constant for all $\lambda_0\ge 0$.
    
    Combining (i)-(iii) completes the proof.
\end{IEEEproof}

\begin{proposition}
    \label{alphascale_two}
    In two-user case, the feasible interval of $\bm{\alpha}$ can be obtained as
    \begin{equation}
        \label{feasibleinterval_two}
        \begin{aligned}
            & \underline{\bm{\alpha}_1} = \max \left\{\bm{\alpha}_1 \geq \bm{0}\mid\|({\bm{\phi}^{1}_{\text{I}}})^{H}\bm{\alpha}_1\|^{2} \geq \gamma_{\min}\sigma^{2}, \|{\bm{\phi}^{1}_{\text{E}}}\bm{\alpha}_1\|^{2}\geq P^{\min}/\zeta_1\right\},\\
            & \overline{\bm{\alpha}_1}=\left\{\bm{\alpha}_1\geq \bm{0}\mid\|\bm{\alpha}_1\|_2\leq 1 \right\},
        \end{aligned}
    \end{equation}
where $e^1_{\text{I}}(\bm{\alpha}) \triangleq \mathbf{h}_1^H \mathbf{G}\bm{\Lambda}(\bm{\alpha})\mathbf{w}_1 \triangleq ({\bm{\phi}^{1}_{\text{I}}})^H \bm{\alpha}$ and $e^1_{\text{E}}(\bm{\alpha}) \triangleq {\mathbf{h}'}_1^H \mathbf{G}\bm{\Lambda}(\bm{\alpha})\mathbf{w}_1 \triangleq ({\bm{\phi}^{1}_{\text{E}}})^H \bm{\alpha}$, ${\bm{\phi}^{1}_{\text{I}}}$ is the coefficient vector for the linear form of $e^1_{\text{I}}$ about $\bm{\alpha}$ and ${\bm{\phi}^{1}_{\text{E}}}$ is the coefficient vector for the linear form of $e^1_{\text{E}}$ about $\bm{\alpha}$.
\end{proposition}

\begin{IEEEproof}
    From \eqref{SINRconstraint}, we can get $|\mathbf{h}_1^H \mathbf{G}\bm{\Lambda}(\bm{\alpha}_1)\mathbf{w}_1|^2 \ge \gamma_{\min} \sigma^{2}$, which can be transformed into $\|({\bm{\phi}^{1}_{\text{I}}})^{H}\bm{\alpha}_1\|^{2} \geq \gamma_{\min}\sigma^{2}$. From \eqref{powerharvest_low}, $\|{\bm{\phi}^{1}_{\text{E}}}\bm{\alpha}_1\|^{2}\geq P^{\min}/\zeta_1$ can be derived. From \eqref{powerconstraints1}, $\|\bm{\alpha}_1\|_2\leq 1$ is obtained. Thus, the lower bound and upper bound of $\bm{\alpha}_1$ can be derived as \eqref{feasibleinterval_two}. 
\end{IEEEproof}

It is worth noting that due to the strong coupling between the beamforming design and the spatial configuration of the pinching antennas, as well as the presence of nonlinear energy harvesting constraints, the resulting optimization problem is both highly non-convex and NP-hard. The fractional structure of the PCE objective further complicates the problem, making conventional convex optimization approaches inapplicable. By exploiting the AO and Dinkelbach approach, the proposed algorithm is capable of efficiently handling the non-convexity and providing a high-quality solution to the original PCE maximization problem for two-user case.

\subsection{Lower-level Problem Solution of Sum Rate Maximization}

\begin{algorithm}[!t]
    \caption{WMMSE-Based Sum Rate Maximization Algorithm in PASS-WPT for Two-User Case}
    \label{alg:two_sumrate}
    \small
    \begin{algorithmic}[1]
    \REQUIRE Fixed $\mathbf{w}^*,\,\mathbf{X}^*$ and feasible interval $\underline{\bm{\alpha}_1} \le \bm{\alpha} \le \overline{\bm{\alpha}_1}$, $T$, $\epsilon_1$
    \ENSURE $\bm{\alpha}_1^*$
    \STATE Based on \textbf{Algorithm \ref{alg:two_pce}}
    \STATE \textbf{Initialize} feasible $\bm{\alpha}^{(0)}$ with $\|\bm{\alpha}^{(0)}_n\|_2\!\le\!1$, set $T\!\leftarrow\!0$
    \REPEAT
        \STATE \textbf{Build Linear Forms:}
        $e^1_{\text{I}}(\bm{\alpha})$, $e^1_{\text{E}}(\bm{\alpha})$
        \STATE \textbf{WMMSE Receive Filter:} $u_1$
        \STATE \textbf{Transform as SOCP}: (\ref{min_sumrate11})
        \STATE $\bm{\alpha}^{(T+1)} \in \arg\min_{\bm{\alpha}}~ \bm{\alpha}^H\bm{\phi}^1_{\text{I}}(\bm{\phi}^1_{\text{I}})^H\bm{\alpha}-2\,\Re\{u_1\bm{\phi}^1_{\text{I}}\bm{\alpha}\}$
        \STATE $\text{\textbf{s.t.}}
        \quad |(\bm{\phi}_{\text{E}}^1)^H\bm{\alpha}|\!\ge\!\sqrt{\Gamma},\ 
        \|\bm{\alpha}_n\|_2\!\le\!1,~
        \bm{\alpha}\!\ge\!\mathbf{0}$
        \STATE \textbf{KKT: Phase Fixing for SOC}
        \STATE Set $\Theta=\angle\big((\bm{\phi}_{\text{E}}^1)^H\bm{\alpha}^{(T)}\big)$, replace $|(\bm{\phi}_{\text{E}}^1)^H\bm{\alpha}|\!\ge\!\sqrt{\Gamma}$ by 
        $\Re\{e^{-j\Theta}(\bm{\phi}_{\text{E}}^1)^H\bm{\alpha}\}\!\ge\!\sqrt{\Gamma},\ \Im\{e^{-j\Theta}(\bm{\phi}_{\text{E}}^1)^H\bm{\alpha}\}\!=\!0$
        \STATE \textbf{Update} $T\!\leftarrow\!T+1$
    \UNTIL{$\big|\log_2(1+|\bm{\phi}^1_{\text{I}}|^2/\sigma^2)\big| < \epsilon_1$}
    \STATE \textbf{RETURN} $\bm{\alpha}_1^*=\bm{\alpha}^{(T)}$
    \end{algorithmic}
    \end{algorithm}

In the two-user case with one IDR and one EHR, the sum rate maximization problem reduces to maximizing the single IDR's achievable rate subject to the EHR's energy harvesting constraint. The IDR's SINR degenerates to an SNR since no multi-user interference exists. The EHR does not contribute to the rate but imposes a minimum harvested power requirement. 
Based on the above optimized $\mathbf{w}_1^*$, $\mathbf{X}^*_{\text{two}}$, and feasible interval $\underline{\bm{\alpha}_1} \le \bm{\alpha}_1 \le \overline{\bm{\alpha}_1}$ for two-user case, we can transform the sum rate maximization problem into data rate maximization problem to finetune $\bm{\alpha}_1$ as follows, 
\begin{subequations}
    \label{max_sumrate10}
    \begin{align}
            \mathbf{P}_{1.0}: \quad & \max_{\bm{\alpha}}\log_2 \left(1+\mathrm{SNR}_1(\mathbf{w}_1,\mathbf{X})\right),
            \label{maxsumrate2}\\
            \mathrm{s.t.~} \quad & R_{1} = \log_2 \left(1+\mathrm{SNR}_1(\mathbf{w}_1,\mathbf{X})\right) \ge R_{1}^{\mathrm{min}},k=K=1,
            \label{rateconstraints1}\\
            & \text{(\ref{alphaconstraint})},
    \end{align}
\end{subequations}
where the $\mathrm{SNR}_1$ represents signal-to-noise ratio (SNR) which ignores the multi-user interference. We notice that the system sum rate is reduced to the data rate because of the only one IDR, expressed as
\begin{equation}
    \label{SNR}
    \mathrm{SNR}_1(\bm{\alpha})=\frac{\left|\mathbf{h}_1^H \mathbf{G}\bm{\Lambda}(\bm{\alpha})\mathbf{w}_1\right|^2}{\sigma^2},
\end{equation}
where this expression involves the multiplicative coupling among the transmit beamforming $\mathbf{w}_1$, the PA position matrix $\mathbf{X}$, and the power radiation radio $\bm{\alpha}$. Hence, $\mathbf{P}_{1.0}$ is a highly non-convex optimization problem. Based on the optimized transmit beamforming $\mathbf{w}^*_1$ and PA position matrix $\mathbf{X}^*_{\text{two}}$, the data rate maximization problem $\mathbf{P}_{1.0}$ can be reduced to only optimize the power radiation ratio matrix. As for $\mathrm{SNR}_1$, $\bm{\alpha}$ is quadratic form in this expression resulting in non-convexity of $\mathrm{SNR}_1$. 
To derive the optimized solution of $\mathbf{P}_{1.0}$, we utilize WMMSE technology \cite{shi2011wmmse} to transform the original non-convex maximization problem into convex optimization subproblem with the auxiliary variable $e^1_{\text{I}}(\bm{\alpha})$.
Let the estimated signal vector of the IDR be $\hat{s}_1 = u_1 y_1$, denoted by the linear receive filter $u_1$ for the IDR, $u_1 = \frac{e^1_{\text{I}}}{|e^1_{\text{I}}|^2+\sigma^2}$.

At WMMSE equivalent, (\ref{maxsumrate2}) is equivalent to minimizing the following convex secondary targets
\begin{equation}
    \label{min_convexalpha}
    \min_{\bm{\alpha}} \quad|u_1|^{2}\bm{\alpha}^{H}\underbrace{({\bm{\phi}^{1}_{\text{I}}} ({\bm{\phi}^{1}_{\text{I}}})^H)}_{{\bm{\phi}^{1}_{\text{I}}} ({\bm{\phi}^{1}_{\text{I}}})^H \succeq 0}\bm{\alpha}\mathrm{~-~2~}\Re\{u_1\bm{\alpha}^{H}{\bm{\phi}^{1}_{\text{I}}}\}.
\end{equation}

Furthermore, we can rewrite this minimization problem as the convex quadratically constraint quadratic programming (QCQP) form,
\begin{subequations}
    \label{min_sumrate11}
    \begin{align}
            \mathbf{P}_{1.1}: \quad & \min_{\bm{\alpha}} \quad|u_1|^{2}\bm{\alpha}^{H}\underbrace{({\bm{\phi}^{1}_{\text{I}}} ({\bm{\phi}^{1}_{\text{I}}})^H)}_{{\bm{\phi}^{1}_{\text{I}}} ({\bm{\phi}^{1}_{\text{I}}})^H \succeq 0}\bm{\alpha}\mathrm{~-~2~}\Re\{u_1\bm{\alpha}^{H}{\bm{\phi}^{1}_{\text{I}}}\},
            \label{min_QCQP}\\
            \mathrm{s.t.~} \quad & \left|({\bm{\phi}^{1}_{\text{E}}})^H \bm{\alpha}\right| \ge \sqrt{\frac{p_0}{\zeta_1\|\mathbf{w}_1\|^2}},
            \label{rateforSOCP}\\
            & \eqref{feasibleinterval_two}
    \end{align}
\end{subequations}
where (\ref{rateforSOCP}) is converted by (\ref{rateconstraints1}). We can see that $\mathbf{P}_{1.1}$ is a standard second-order core programming (SOCP), and hence, it can be solved by interior point methods, i.e., Karush-Kuhn-Tucker (KKT) \cite{lange2012kkt}. Let $\lambda_n$ be the two-norm cone multiplier of the waveguide $n$, $\eta_1$ and $\mu_1$ be non-negative constrained multipliers (element by element). Then, the complementary relaxation can be expressed as
\begin{equation}
    \label{relaxation_lambda1}
    \begin{aligned}
        & \lambda_{n}\left(\|\bm{\alpha}_n^*\|_{2}-1\right)=0,\\
        & \eta_1{\left(\Re\{e^{-j\theta}{(\bm{\phi}^1_{\text{E}})}^{H}\bm{\alpha}_n^*\}-\sqrt{\Gamma}\right)}=0,
    \end{aligned}
\end{equation}
and original feasibility is formulated as $\mu_{1}\bm{\alpha}_n^*=0$.

Finally, we can obtain the optimal solution of this minimization problem, that is the optimal power radiation ratios
\begin{equation}
    \label{alpha_1_*}
    \bm{\alpha}^*_1 = [{\bm{\phi}^{1}_{\text{I}}} ({\bm{\phi}^{1}_{\text{I}}})^H]^{+}(u_1\bm{\phi}^1_{\text{I}} + \eta^*_1 e^{-j\theta} \bm{\phi}^1_{\text{E}}),
\end{equation}
where $\eta^*_1$ is determined by $\Re\{e^{-j\theta}{(\bm{\phi}^1_{\text{E}})}^{H}\bm{\alpha}_1^*\} = \sqrt{\frac{p_0}{\zeta_1\|\mathbf{w}^*_1\|^2}}$.


\subsection{Computational Complexity Analysis}

The joint optimization algorithms for the two-user case are summarized in Algorithm \ref{alg:two_pce} and \ref{alg:two_sumrate}. For Algorithm \ref{alg:two_pce}, the Dinkelbach outer loop is executed by $T_{\text{OD}}$ iterations to derive the transmit beamforming on two-dimensional space, which costs $\mathcal{O}(NL)$. The inner loop is based on AO to optimize the PA positions, which requires a gradient and projection of $\mathcal{O}(NL)$. It is noticed that $\bm{\alpha}_n$ is iteratively updated to maximize $\text{PCE}$ avoiding stuck in local optimum, and hence, this part is updated by interior point method with the cost of $\mathcal{O}(T_{\text{OD}} T_{\text{IA}}({NL})^3)$, where $T_{\text{IA}}$ is the required iterations of interior point method. Thus, the computational complexity of Algorithm \ref{alg:two_pce} is $\mathcal{O}(NL)+\mathcal{O}(T_{\text{OD}} T_{\text{IA}}({NL})^3)$.

For Algorithm \ref{alg:two_sumrate}, each WMMSE round updates a scalar equalizer in $\mathcal{O}(1)$ (with $T_{\text{OW}}$ iterations) and then solves the SOCP of the same size with the above $NL$. Consequently, the computational complexity is dominated by the SOCP and is given by $\mathcal{O}(T_{\text{OW}} T_{\text{IA}}({NL})^3)$.

In summary, the computational burden is dominated by solving the optimal power radiation ratios with decision dimension $NL$ for the two-user case. The beamforming and position updates are linear time in $NL$, and thus, do not affect the asymptotic order.

\section{Optimization Of PASS-WPT For Multi-User Case}

In this section, we consider the multi-user case of the PASS-WPT, which consists of $K$ IDRs and $Q$ EHRs served by PAs along $N$ waveguides. It is noticed that the AO and WMMSE-based algorithms are suitable for two-user case, because two-user scenario has a low interference dimension, $\mathbf{w}_1$ falls in a two-dimensional tensor space and can be solved in closed-form with a one-dimensional proportional parameter. For each discrete PA position are convex, so that an exhaustive enumeration of all PA positions yields a closed-form solution for the discrete domain at low computational cost.

In contrast, the multi-user situation exists $\mathcal{O}(K^2Q)$ level interference and multi-constraint coupling. If AO and WMMSE-based algorithms are used separately, the two sets of cores and nested external loops will lead to a rapid increase in value and implementation complexity as the scale increases. Hence, we design a unified approach based on quadratic transform-Lagrangian dual transform \cite{kesten1970QT} to meanwhile solve PCE and sum rate maximization. Specifically, through closed-form auxiliary variables we construct tight low bounds to convex the $\mathbf{X}$ update step for $\mathbf{W}^{\star}$, $\mathbf{X}^{\star}$ and $[\underline{\bm{\alpha}}, \overline{\bm{\alpha}}]$, and $\bm{\alpha}$ update step for refining $\bm{\alpha}^{\star}$ as QCQP/SOCP and to guarantee monotonic convergence.

We develop a unified algorithm that handles both (\ref{PCE0}) and (\ref{max_sumrate1}) through tight surrogate objectives constructed via QT-LDT. Given a fixed $\mathbf{X}$, each outer iteration performs the sequence that updates auxiliary variables, and then, executes convex $\mathbf{X}$ update step, finally, solves the convex $\bm{\alpha}$.
The solution procedures of $\mathbf{P}_0$ and $\mathbf{P}_1$ differ only in the auxiliary variables and coefficients, while both yield monotone ascent and converge to a KKT point for the fixed $\mathbf{X}$.

\subsection{For PCE Maximization}

Define the EHR side effective channel $\mathbf{d}^H_q\triangleq \mathbf{h}'_q(\mathbf{X})^{H}\mathbf{G}(\mathbf{X})\bm{\Lambda}(\bm{\alpha})$ with consistency of \eqref{c_d}, and stack $v_{qk}\triangleq \sqrt{\zeta_q}\,\mathbf{d}_q^{H}\mathbf{w}_k$ into $\mathbf{v}\in\mathbb{C}^{QK}$, so the numerator in \eqref{PCE0} becomes
$\|\mathbf{v}\|_2^2$ while the denominator is $D_k(\mathbf{W})=\phi\sum_k\|\mathbf{w}_k\|_2^2+Q\,P_C$.

\begin{lemma}
\label{lem:pce-tight}
For any $\|\mathbf{u}_{\rm E}\|_2\le 1$ and $\rho\in\mathbb{C}$, the tight lower bound of \eqref{PCE0} can be written as 
\begin{equation}
    \label{PCE-sur}
    \begin{aligned}
        \frac{\|\mathbf{v}\|_2^2}{D_k} \ & =\ \max_{\|\mathbf{u}_E\|_2\le 1, \ \rho\in\mathbb{C}}\ \Big[\,2\,\Re\{\rho^{*}\mathbf{u}_{\rm E}^{H}\mathbf{v}\}-|\rho|^2\,d\,\Big], \\
        \mathbf{u}_{\rm E}^{\star} & = \frac{\mathbf{v}}{\|\mathbf{v}\|_2}\ (\mathbf{v}\ne\mathbf{0}),\ \rho^{\star}=\frac{\mathbf{u}_E^{H}\mathbf{v}}{D_k}.
    \end{aligned}
\end{equation}
\end{lemma}
\begin{IEEEproof}
By Cauchy-Schwarz, $\max_{\|\mathbf{u}_E\|\le 1}\Re\{\mathbf{u}_{\rm E}^{H}\mathbf{v}\}=\|\mathbf{v}\|_2$ achieved at $\mathbf{u}_{\rm E}=\mathbf{v}/\|\mathbf{v}\|_2$ if $\mathbf{v}\!\ne\!\mathbf{0}$.
For fixed $\mathbf{u}_{\rm E}$, maximize over $\rho$ the concave quadratic
$2\Re\{\rho^{*}\mathbf{u}_{\rm E}^{H}\mathbf{v}\}-|\rho|^2 d$, whose optimum is $\rho=(\mathbf{u}_{\rm E}^{H}\mathbf{v})/{D_k}$ with value 
$|\mathbf{u}_{\rm E}^{H}\mathbf{v}|^2/{D_k}$. The outer maximization over $\|\mathbf{u}_{\rm E}\|\le 1$ then gives $\|\mathbf{v}\|_2^2/{D_k}$, with equality at \eqref{PCE-sur}.
\end{IEEEproof}

With the surrogate in \eqref{PCE-sur} tight at the current iterate, we can process $\mathbf{W}$ and $\bm{\alpha}$ update steps.

\subsubsection{$\mathbf{X}$ update step}
For fixed $(\mathbf{u}_{\rm E},\rho,\boldsymbol{\alpha})$,
\begin{subequations}
    \begin{align}
        \nonumber
        \max_{\mathbf{W},\mathbf{X},\underline{\bm{\alpha}}, \overline{\bm{\alpha}}}\  & 2\,\Re\!\Big\{\sum_{q,k}\sqrt{\zeta_q}\,\rho^{*}u_{{\rm E},qk}^{*}\,\mathbf{d}_q(\mathbf{X},\bm{\alpha})^{H}\mathbf{w}_k\Big\}\\
        & -|\rho|^2\,\phi\sum_{k}\|\mathbf{w}_k\|_2^2 
        \label{eq:W-pce-III}\\
        \text{s.t.}\  & \eqref{transmitpower_high},
        \end{align}
\end{subequations}
this is a convex SOCP, and thus, it can be solved by the interior point methods as above.

\subsubsection{$\bm{\alpha}$ update step}
Based on the above, we solve
\begin{subequations}
    \begin{align}
        \max_{\bm{\alpha}}\ \
        &2\,\Re\!\Big\{\sum_{q,k}\sqrt{\zeta_q}\,\rho^{*}u_{{\rm E},qk}^{*}\,
        \mathbf{h}'_q(\mathbf{X})^{H}\mathbf{D}_E(\mathbf{w}_k,\mathbf{X})\,\boldsymbol{\alpha}\Big\}
        \label{alpha-pce}\\
        \text{s.t.}\ \
        &\eqref{rateconstraints}-\eqref{alphaconstraint}.
    \end{align}    
\end{subequations}

It can be observed that (\ref{alpha-pce}) is a linear objective with SOC constraints (\ref{rateconstraints})-(\ref{alphaconstraint}), i.e., a single SOCP.

During the $\mathbf{X}$ update steps and $\bm{\alpha}$ update steps, the auxiliary variables can be ultimately updated in closed-form,
\begin{equation}
    \label{PCE-aux}
    \mathbf{u}^{\star}_{\rm E} = \frac{\mathbf{v}}{\max\{\|\mathbf{v}\|_2,\epsilon\}},
    \quad 
    \rho^{\star} = \frac{\mathbf{u}_{\rm E}^{H}\mathbf{v}}{\phi\sum_{k}\|\mathbf{w}_k\|_2^2+Q\,P_{\text{C}}}, \ \ \epsilon>0.
\end{equation}

\begin{proposition}
\label{prop:pce-mono}
At iteration $t$, the surrogate in \eqref{PCE-sur} is tight at $(\mathbf{W}^{(t)},\bm{\alpha}^{(t)})$ and is lower-bound of the true PCE. Maximizing it over $(\mathbf{W},\bm{\alpha})$ produces a non-decreasing PCE sequence converging to a KKT point at the discrete $\mathbf{X}$.
\end{proposition}

\begin{IEEEproof}
Follows directly from Lemma~\ref{lem:pce-tight}, similar proof with Proposition~\ref{prop:sr-mono}. The KKT conditions with the constraints \eqref{transmitpower_high} (multiplier $\mu\!\ge\!0$) give the closed-form expression, 
\begin{equation}
    \label{eq:w_star_PCE}
    \mathbf{W}^{\star}=\frac{\sqrt{\zeta_q}\mathbf{h}'_q(\mathbf{X}^{\star})^{H}\mathbf{G}(\mathbf{X}^{\star})\bm{\Lambda}}{\,|\rho^{\star}|^2\,\phi+\mu^{\star}\,},
\end{equation}
with $\mu$ chosen to satisfy the power budget:
\begin{equation}
    \label{eq:mu_star_PCE}
    \begin{aligned}
        \mu^{\star}=
        \begin{cases}
            0, \displaystyle \qquad \frac{\|\sqrt{\zeta_q}\mathbf{h}'_q(\mathbf{X})^{H}\mathbf{G}(\mathbf{X}^{\star})\bm{\Lambda}\|_F^2}{(|\rho|^2\phi)^2} \le  P^{\max}, & \\
            \frac{\|\sqrt{\zeta_q}\mathbf{h}'_q(\mathbf{X})^{H}\mathbf{G}(\mathbf{X}^{\star})\bm{\Lambda}\|_F}{\sqrt{P^{\max}}}-|\rho|^2\phi, \displaystyle \ \ \text{otherwise}, &
        \end{cases}
    \end{aligned}
\end{equation}
and optimized PA position matrix can be derived as
\begin{equation}
    \label{optimal_x}
    \mathbf{X}^{\star} \in \arg \max_{\mathbf{X}\in \mathcal{X}} \dfrac{\sum_{q}\zeta_q\sum_{k}\big|\mathbf{h}'_q{}^{H}\mathbf{G}\bm{\Lambda}\mathbf{w}_k\big|^2}{\phi\sum_{k}\|\mathbf{w}_k\|_2^2+Q P_{\text{C}}}.
\end{equation}

\end{IEEEproof}

\begin{proposition}
    \label{feasibleinterval_alpha}
    For multi-user case, the feasibility interval of $\bm{\alpha}$ can be derived as
    \begin{equation}
        \label{feasibleinterval_mul}
        \begin{aligned}
            \underline{\bm{\alpha}}(\mathbf{W}^{\star},\mathbf{X}^{\star}) = & \{ \bm{\alpha} \ge \mathbf{0} | \bm{\alpha}^H [\mathbf{h}_k^{H}\bm{\beta}-\gamma_{\min}\sum_{j\neq k}\mathbf{h}_k^{H}\bm{\beta}]\bm{\alpha}\ge \gamma_{\min} \sigma^2, \\
            & \|\mathbf{d}^H_q \bm{\alpha}\|^2 \ge P^{\min}/\zeta_q, \forall q \in \mathcal{Q}, \forall k \in \mathcal{K}\},\\
            \overline{\bm{\alpha}}(\mathbf{W}^{\star},\mathbf{X}^{\star}) = & \{\bm{\alpha} \ge \mathbf{0} | \|\bm{\alpha}\|^2 \le 1, \forall q \in \mathcal{Q}, \forall k \in \mathcal{K}\}.
        \end{aligned}
    \end{equation}
\end{proposition}
\begin{IEEEproof}
    The proof process is similar with Proposition \ref{alphascale_two}.
\end{IEEEproof}

\subsection{For Sum Rate Maximization}

We introduce $\gamma_k = \mathrm{SINR}_k \in \mathbb{R}$ and $\nu_k \in \mathbb{C}$. We first operate with natural logarithms and convert to $\log_2(\cdot)$ via $1/\ln 2$ afterwards. Let the lower bound function of data rate $R_k$ of IDR $k$ as $\hat{\Phi}_k = \Phi_k / \ln 2$, 
\begin{equation}
    \label{lowerboundf_rk}
    \begin{aligned}
        \Phi_k(\mathbf{W},\bm{\alpha};\gamma_k,\nu_k) = & \ln(1+\gamma_k)-\gamma_k +2\,\Re\!\left\{\sqrt{1+\gamma_k}\,\nu_k\,s_{k,k}\right\} \\
        & -|\nu_k|^2\!\left(\sum_{j=1}^{K}|s_{k,j}|^2+\sigma^2\right),
    \end{aligned}
\end{equation}
where $s_{k,j}(\mathbf{W}, \mathbf{X},\bm{\alpha}) \triangleq  \mathbf{h}^{H}_k(\mathbf{X})\,\mathbf{G}(\mathbf{X})\,\bm{\Lambda}(\bm{\alpha})\,\mathbf{w}_j$.


\begin{lemma}{Tight lower bound for $R_k$:}
\label{lem:tight}
For any $\gamma_k\ge 0$ and $\nu_k\in\mathbb{C}$, $R_k = \log_2(1+{\rm SINR}_k)\ge \Phi_k(\cdot)/\ln2$ with equality at
\begin{equation}
    \label{eq:sr-aux-opt}
    \gamma_k^{\star}=\frac{|s_{k,k}|^2}{\sum_{j\ne k}|s_{k,j}|^2+\sigma^2},\quad
    \nu_k^{\star}=\frac{\sqrt{1+\gamma_k^{\star}}\,s_{k,k}}{\sum_{j=1}^{K}|s_{k,j}|^2+\sigma^2}.
\end{equation}
\end{lemma}
\begin{IEEEproof}
Let $D_k\triangleq\sum_{j}|s_{k,j}|^2+\sigma^2$ and $S_k\triangleq |s_{k,k}|^2$.
With the fixed $\gamma_k$, $\Phi_k$ is a concave quadratic in $\nu_k$, rewritten (\ref{lowerboundf_rk}):
\begin{equation}
    \label{lowerboundf_rk_simply}
    \Phi_k=\ln(1+\gamma_k)-\gamma_k+2\Re\{\sqrt{1+\gamma_k}\,\nu_k s_{k,k}\}-D_k|\nu_k|^2.
\end{equation}

Maximizing over $\nu_k$ by completing the square yields
$\nu_k^{\rm opt}=\frac{\sqrt{1+\gamma_k}\,s_{k,k}}{D_k}$ and $\max_{\nu_k} \Phi_k (\nu_k,\gamma_k)= \max_{\nu_k} \ln(1+\gamma_k)-\gamma_k+\frac{(1+\gamma_k)S_k}{D_k}$.
For simplicity, maximizing $f(\gamma_k)\triangleq \ln(1+\gamma_k)-\gamma_k+\dfrac{(1+\gamma_k)S_k}{D_k}$ over $\gamma_k\ge 0$ by the derivation of $\Phi_k$: 
\begin{equation}
    \label{derivative_Phik}
    {\Phi}^{'}_k (\gamma_k)=\frac{1}{1+\gamma_k}-1+\frac{S_k}{D_k} = -\frac{\gamma_k}{1+\gamma_k}+\frac{S_k}{D_k}.
\end{equation}

Setting $\Phi^{'}_k(\gamma_k)=0$ gives $\dfrac{\gamma_k}{1+\gamma_k}=\dfrac{S_k}{D_k}$,
hence $\gamma_k^{\star}=\dfrac{S_k}{D_k-S_k}=\dfrac{|s_{k,k}|^2}{\sum_{j\ne k}|s_{k,j}|^2+\sigma^2}$.
Substituting $\gamma_k^{\star}$ and $\nu_k^{\star}$ into $\Phi_k$ gives
$\max_{\gamma_k,\nu_k}\Phi_k=\ln\!\big(1+\dfrac{S_k}{D_k-S_k}\big)=\ln(1+{\rm SINR}_k)$, establishing tightness and \eqref{eq:sr-aux-opt}.
\end{IEEEproof}

Denote $\mathcal{L}_{\rm SR}(\mathbf{W},\boldsymbol{\alpha};\boldsymbol{\gamma},\boldsymbol{\nu})\triangleq \sum_{k}\Phi_k$.
At iteration $t \in \mathcal{T}$, it is targeted to maximize $\mathcal{L}_{\rm SR}$ over $(\mathbf{W},\bm{\alpha})$ through the $\mathbf{X}$ update step and $\bm{\alpha}$ update step that $(\boldsymbol{\gamma},\boldsymbol{\nu})$ is updated by \eqref{eq:sr-aux-opt} using $(\mathbf{W}^{(t)},\boldsymbol{\alpha}^{(t)})$ until the closed-form $(\mathbf{W}^{*},\boldsymbol{\alpha}^{*})$. The $\mathbf{X}$ update step and $\bm{\alpha}$ update step are shown as follows:

\subsubsection{$\mathbf{X}$ update step}
Using $s_{k,j}=\mathbf{h}_k^{H}\mathbf{G}\bm{\Lambda}\mathbf{w}_j$ and collecting the $\mathbf{W}$-terms,
\begin{subequations}
    \begin{align}
        \nonumber
        \max_{\mathbf{W},\mathbf{X},\underline{\bm{\alpha}}, \overline{\bm{\alpha}}} & \ \sum_{k}2\,\Re\!\left\{\sqrt{1+\gamma_k}\,\nu_k^{\!*}\,\mathbf{h}^H_k(\mathbf{X})\mathbf{G}(\mathbf{X})\bm{\Lambda}\mathbf{w}_k\right\} \\
        & \ -\sum_{k}\mathbf{w}^H_k\!\Big(\sum_{j}|\nu_j|^2\bm{\Lambda}^H\mathbf{G}^H\mathbf{h}_j\mathbf{h}_j^{H}\mathbf{G}\bm{\Lambda}\Big)\mathbf{w}_k
        \label{max_W_sumrate}\\
        \mathrm{s.t.~} & \quad \text{(\ref{rateconstraints})-(\ref{alphaconstraint})}.
    \end{align}
\end{subequations}

It can be observed that (\ref{max_W_sumrate}) is a convex SOCP, which can be solved by interior point method (similar process as (\ref{min_QCQP})-(\ref{alpha_1_*})). Additionally, the constraints (\ref{rateconstraints})-(\ref{alphaconstraint}) are SOC forms without destroying the convexity.

\subsubsection{$\bm{\alpha}$ update step}

Denote $s_{k,j}=\mathbf{h}_k^{H}\mathbf{G}\bm{\Lambda}\mathbf{w}_j = \mathbf{h}_k^{H}\bm{\beta}(\mathbf{w}_j,\mathbf{X})\,\bm{\alpha}$ with
$\bm{\beta}(\mathbf{w}_j,\mathbf{X})\triangleq \mathrm{diag}(\mathbf{G}(\mathbf{X})\odot \mathbf{w}_j)$, and thus the subproblem of $\boldsymbol{\alpha}$ becomes
\begin{subequations}
    \begin{align}
        \nonumber
        \max_{\boldsymbol{\alpha}}\ \ 
        &\sum_{k}2\,\Re\!\left\{\sqrt{1+\gamma_k}\,\nu_k^{\!*}\,\mathbf{h}_k(\mathbf{X})^{H}\bm{\beta}(\mathbf{w}_k,\mathbf{X})\,\boldsymbol{\alpha}\right\} \\
        & -\boldsymbol{\alpha}^{H}\!\Big(\sum_{k}\sum_{j}|\nu_k|^2\,\bm{\beta}(\mathbf{w}_j,\mathbf{X})^{H}\mathbf{h}_k\mathbf{h}_k^{H}\bm{\beta}(\mathbf{w}_j,\mathbf{X})\Big)\boldsymbol{\alpha}
        \label{eq:alpha-sr}\\
        \text{s.t.}\ \ \ \
        & \eqref{rateconstraints}-\eqref{alphaconstraint},
    \end{align}
\end{subequations}
where \eqref{eq:alpha-sr} is convex QCQP.
Let the PSD quadratic form be $\mathbf{B}_{\rm SR}=\sum_k \sum_{j}|\nu_k|^2\,\bm{\beta}(\mathbf{w}_j,\mathbf{X})^{H}\mathbf{h}_k\mathbf{h}_k^{H}\bm{\beta}(\mathbf{w}_j,\mathbf{X}) \succeq\mathbf{0}$. For simplicity, we define $\mathbf{B}_{\rm SR}=\mathbf{R}^H\mathbf{R}$.

With $\mathbf{b}=\mathbf{R}^{-H}\!\sum_k \sqrt{1+\gamma_k}\,\nu_k^{\!*}\bm{\beta}(\mathbf{w}_k,\mathbf{X})^{H}\mathbf{h}_k$, \eqref{eq:alpha-sr} is equivalent to the SOCP through characteristic decomposition of $\mathbf{B}_{\rm SR}$, which can be formulated as
\begin{subequations}
    \begin{align}
        \min_{\boldsymbol{\alpha},\,b_{\rm SR}}\ \ & \tfrac12 {b^2_{\rm SR}}-\tfrac12\|\mathbf{b}\|_2^2
        \label{eq:alpha-sr-socp}\\
        \text{s.t.}\ \ \ \ & \|\mathbf{R}\boldsymbol{\alpha}-\mathbf{b}\|_2\le b_{\rm SR},
        \label{alpha_SOCPconstraint}\\
        & \eqref{rateconstraints}-\eqref{alphaconstraint}.
    \end{align}
\end{subequations}

During the $\mathbf{X}$ update steps and $\bm{\alpha}$ update steps, the auxiliary variables can be ultimately updated in closed-form $\gamma^{\star}_k$, $\nu^{\star}_k$.
Given the updated $(\mathbf{X},\bm{\alpha})$, compute $s_{k,j}$ and \eqref{eq:sr-aux-opt}.

Since each iteration tightens the lower bound, and then, maximizes the lower bound, the original sum rate monotonically does not decrease and converges to the KKT point at the discrete $\mathbf{X}$.

\begin{proposition}
\label{prop:sr-mono}
At iteration $t$, $\mathcal{L}_{\rm SR}$ is tight at $(\mathbf{W}^{(t)},\bm{\alpha}^{(t)})$ and serves as a lower bound on the true sum rate. Maximizing $\mathcal{L}_{\rm SR}$ over $(\mathbf{W},\bm{\alpha})$ yields $\sum_k\log_2(1+{\rm SINR}_k)$ non-decreasing in $t$ and convergence to a KKT point of the sum rate maximization at the discrete $\mathbf{X}$. When $\mathbf{X}^{\star}$ is searched, $\bm{\alpha}^{\star}$ can be derived as the closed-form expression.
\end{proposition}

\begin{IEEEproof}
By Lemma~\ref{lem:tight} and the construction of \eqref{lowerboundf_rk}, 
$\mathcal{L}_{\rm SR}$ equals the true value of sum rate (up to $1/\ln 2$) at the current iteration and is a global lower bound. Moreover, maximizing it cannot decrease the original objective and finally converge to a stationary KKT point.
Its KKT closed-form expression is
    \begin{equation}
    \label{eq:alpha_star_KKT}
    \big(\mathbf{B}_{\rm SR} + \sum_{n}\lambda_n\,\mathbf{P}_n + \mathbf{L}\big)\ \bm{\alpha}^{\star}
    \ =\ \mathbf{b},
    \lambda_n\ge 0,\ \ \mathbf{L}={\rm diag}(\mathbf{\bm{\iota}})\ge \mathbf{0},
    \end{equation}
    together with the complementarity conditions
    \begin{equation}
    \label{eq:alpha_KKT_comp}
    \begin{cases}
        \lambda_n\big(\|\bm{\alpha}_n^{\star}\|_2-1\big)=0,\  \forall n,\\
        \iota_{n,l}\,\alpha_{n,l}^{\star}=0,\ \ \quad \qquad \forall l,\\
        \|\bm{\alpha}_n^{\star}\|_2\le 1,  \qquad \bm{\alpha}^{\star}\ge \mathbf{0}.
    \end{cases}
    \end{equation}

    Here $\mathbf{P}_n$ is the block-diagonal projector, that is, waveguide selection matrix, that selects the $n$-th waveguide,
    \begin{equation}
        \label{waveguide_p}
        \mathbf{P}_n \!=\! \mathrm{diag}\big(\!\underbrace{\mathbf{0}_{1}}_{\text{waveguide}1},\ldots,
        \underbrace{\mathbf{0}_{n-1}}_{...},
        \underbrace{\mathbf{I}_{n}}_{\text{waveguide }n},
        \underbrace{\mathbf{0}_{n+1}}_{...},\ldots,
        \underbrace{\mathbf{0}_N}_{\text{waveguide}N}\!\big).
    \end{equation}

    The closed-form power radiation ratio is formulated as
    \begin{equation}
        \label{optimal_alpha}
        \begin{aligned}
            \bm{\alpha}^{\star} & =\big(\mathbf{B}_{\rm SR}+\sum_{n}\lambda_n\mathbf{P}_n\big)^{-1}\mathbf{b} \\
            = & (\sum_k \sum_{j}|\nu^{\star}_k|^2\,\bm{\beta}(\mathbf{w}_j,\mathbf{X}^{\star})^{H}\mathbf{h}_k\mathbf{h}_k^{H}\bm{\beta}(\mathbf{w}_j,\mathbf{X}^{\star}) + \sum_{n}\lambda_n\mathbf{P}_n)^{-1}\\
            &  \cdot(\mathbf{R}^{-H}\!\sum_k \sqrt{1+\gamma^{\star}_k}\,\nu_k^{\!\star}\bm{\beta}(\mathbf{w}_k,\mathbf{X}^{\star})^{H}\mathbf{h}_k).
        \end{aligned}
    \end{equation}
\end{IEEEproof}

\begin{algorithm}[!t]
    \caption{QT-LDT-Based Algorithm for Multi-User Case in PASS-WPT}
    \label{alg:multi_user_QTLDT}
    \begin{algorithmic}[1]
    \REQUIRE Discrete $\mathbf{X} \in \mathcal{X}$, $\{\mathbf{h}_k(\mathbf{X})\}_{k=1}^{K}$, $\{\mathbf{h}'_q(\mathbf{X})\}_{q=1}^{Q}$, $\mathbf{G}(\mathbf{X})$, $\bm{\Lambda}(\bm{\alpha})$, $\varepsilon$
    \ENSURE $\mathbf{W}^{\star}$, $\mathbf{X}^{\star}$, $\bm{\alpha}^{\star}$
    \STATE Initialize feasible $(\mathbf{W}^{(0)},\bm{\alpha}^{(0)})$ with $\sum_k\|\mathbf{w}_k^{(0)}\|_2^2\le P_{\max}$, $\|\bm{\alpha}_n^{(0)}\|_2\le 1$, $\bm{\alpha}^{(0)}\ge\mathbf{0}$; set $t = 0$
    \REPEAT
        \IF{$F=\mathrm{PCE}$}
            \STATE Update $\mathbf{u}_{\rm E}^{(t+1)}$, \ \ $\rho^{(t+1)}$ from \eqref{PCE-aux}
        \ELSE
            \STATE Update $\gamma_k^{(t+1)}$, \ \ $\nu_k^{(t+1)}$ from \eqref{eq:sr-aux-opt}
        \ENDIF
            \STATE Solve \eqref{eq:W-pce-III} and \eqref{alpha-pce}
            \STATE Update $\mathbf{W}^{\star} = \mathbf{W}^{(t+1)} = \mathbf{W}^{(t)}$ from \eqref{eq:w_star_PCE}
            \STATE Update $\mathbf{X}^{\star}$ in \eqref{optimal_x} and $\underline{\bm{\alpha}} \le \bm{\alpha} \le \overline{\bm{\alpha}}$ in \eqref{feasibleinterval_mul}
            \STATE Solve \eqref{max_W_sumrate} and \eqref{eq:alpha-sr}
            \STATE Compute $\bm{\beta}(\mathbf{w}_j^{(t+1)},\mathbf{X}) = \mathrm{diag}(\mathbf{G}(\mathbf{X})\odot \mathbf{w}_j^{(t+1)})$, $\mathbf{B}_{\rm SR} = \sum_{k,j}|\nu_k^{(t+1)}|^2\,\bm{\beta}(\mathbf{w}_j)^{H}\mathbf{h}_k\mathbf{h}_k^{H}\bm{\beta}(\mathbf{w}_j)$, $\mathbf{b} = \sum_{k}\sqrt{1+\gamma_k^{(t+1)}}(\nu_k^{(t+1)})^{*}\bm{\beta}(\mathbf{w}_k)^{H}\mathbf{h}_k$
            \STATE Update $\boldsymbol{\alpha}^{(t+1)} = \boldsymbol{\alpha}^{(t)} =\bm{\alpha}^{\star}$ in \eqref{optimal_alpha}
        \STATE $t = t+1$
    \UNTIL Convergence $<\varepsilon$
    \STATE RETURN $\mathbf{W}^{\star}$, $\mathbf{X}^{\star}$, $\bm{\alpha}^{\star}$
    \end{algorithmic}
    \end{algorithm}

\begin{theorem}
\label{thm:KKT}
For either objective (PCE or sum rate), the iteration procedure $\text{auxiliary variables update (tight)}\ \rightarrow\ \mathbf{W} \text{ step (convex)}\ \rightarrow\ \boldsymbol{\alpha} \text{ step (convex)}$ produces a non-decreasing objective sequence bounded above and converges to a KKT point. For the discrete $\mathbf{X}\in\mathcal{X}$ and the attained value $F$ (PCE or sum rate), $({\mathbf{W}^{\star},\mathbf{X}^{\star},\bm{\alpha}^{\star}})\in\arg\max_{\mathbf{X}\in\mathcal{X}}F^{\star}(\mathbf{X})$, 
\begin{equation}
    \label{optimal_wxaplha}
    F^{\star}(\mathbf{X})=\max_{\mathbf{W},\,\mathbf{X},\,\boldsymbol{\alpha}}\ 
    \begin{cases}
        \sum_{k}\log_2\big(1+\mathrm{SINR}_k\big), & \text{(SR)},\\[2pt]
        \dfrac{\sum_{q}\zeta_q\sum_{k}\big|\mathbf{h}'_q{}^{H}\mathbf{G}\bm{\Lambda}\mathbf{w}_k\big|^2}{\phi\sum_{k}\|\mathbf{w}_k\|_2^2+Q P_{\text{C}}}, & \text{(PCE)},
    \end{cases}
\end{equation}
which yields \eqref{eq:w_star_PCE}, \eqref{optimal_x}, \eqref{optimal_alpha}.

\end{theorem}

\begin{IEEEproof}
By Lemma~\ref{lem:tight} or Lemma~\ref{lem:pce-tight}, each surrogate equals the true objective at the $t$-th iteration and monotonically reaches a global lower bound. Since the surrogate attains the true objective at $(\mathbf{W}^{\star},\bm{\alpha}^{\star})$ and shares the same feasible set, classical majorization-minimization (MM) theory ensures KKT convergence since each iteration is solved to closed-form solutions. For each finite $\mathbf{X} \in \mathcal{X}$, the attained value $F$ (SR or PCE) at the convergence point, whose existence is guaranteed by Theorem~\ref{thm:KKT}. Since $\mathcal{X}$ is finite, $({\mathbf{W}^{\star},\mathbf{X}^{\star},\bm{\alpha}^{\star}})\in\arg\max_{\mathbf{X}\in\mathcal{X}}F^{\star}(\mathbf{X})$ is an exact maximization over a finite position set and therefore yields the closed-form solutions. 
\end{IEEEproof}

\subsection{Computational Complexity Analysis}
    Algorithm ~\ref{alg:multi_user_QTLDT} summarizes the QT-LDT algorithm with a discrete PA positions $\mathbf{X} \in \mathcal{X}$. 
    Denote the number of iterations of the outer algorithm by $T_{\rm M}$, the iterations for the scalar $\mu$ in the $\mathbf{X}$ update step by $T_{\mu}$, and the interior-point iterations for $\boldsymbol{\alpha}$ update step by $T_{\rm \gamma}$.
    For PCE maximization, building $\mathbf{d}_q^{H}=\mathbf{h}'_q{}^{H}\mathbf{G}\bm{\Lambda}$ and $v_{qk}=\sqrt{\zeta_q}\,\mathbf{d}_q^{H}\mathbf{w}_k$ to compute the optimized $\mathbf{W}^{\star}$ costs $\mathcal{O}(QK^2+K^2)$. The computation of the optimized PA position $\mathbf{X}^{\star}$ costs $\mathcal{O}(Nj_n)$. The cost of $\bm{\alpha}$ update step is at most $\mathcal{O}(K^2 L^2)$ in the dense worst case because diagonal structure of $\bm{\beta}(\cdot)$ can reduce this in practice. The SOCP solution cost of these can be $\mathcal{O}(K^2 L^2+T_{\rm \gamma} L^3)$. The overall algorithm computational complexity is $\mathcal{O}(TT_{\mu}QK^2Nj_n+TT_{ \mu}K^2Nj_n+TK^2 L^2+TT_{\rm \gamma} L^3)$.


    


\section{Simulation Analysis}

In this simulation, we evaluate the effectiveness of the proposed PASS-WPT framework through numerical simulations.

We evaluate the proposed PASS-WPT transceiver under both two-user and multi-user scenarios. Unless otherwise specified, the carrier frequency is $f_c=28~\mathrm{GHz}$, the effective refractive index of the waveguide is $n_{\rm eff}=1.4$, and the thermal noise power is $\sigma^2=-80~\mathrm{dBm}$. A PASS with $N$ waveguides is deployed at height $h=5~\mathrm{m}$. The two-user case is setting with $\{N,L,K,Q\} = \{1,4,1,1\}$ and the multi-user case is setting with $\{N,L,K,Q\}=\{4,4,4,4\}$. The common minimum requirement of SINR/SNR is denoted by \(\gamma_{\min}\) (typical values $20~\mathrm{dB}$). For two-user case, there is one waveguide pinched along $4$ PAs. When $N=4$, the $y$-axis coordinates of the waveguides are $\{0,10,20,30\}~\mathrm{m}$. Each waveguide pre-installs $L=4$ PAs on a discrete grid along the $x$-axis, with candidate coordinates $\{8,16,24,32\}~\mathrm{m}$ and minimum PA spacing equal to half wavelength at $f_c$. IDRs and EHRs are independently and uniformly dropped in a rectangular region of size $S_x\times S_y$ with $S_x=S_y=10~\mathrm{m}$.

The two-user case is solved by the AO and WMMSE algorithm, whereas the multi-user case is solved by the QT-LDT algorithm. Every convex subproblem is handled by an interior-point solver with relative tolerance \(10^{-4}\); the maximum number of AO rounds is 60, with scaling factor $\epsilon=1.25$.
For discrete PA locations, \(\mathbf{X}\) is selected from the finite grid $\mathcal{X}$ by local-search variant. All curves are averaged over $100$ independent Monte-Carlo drops unless otherwise stated.

We compare the following designs:
\begin{itemize}
    \item \textbf{Conventional MIMO enabled SWIPT}\cite{2013MIMO}: The conventional MIMO framework employs the same number of RF chains with the proposed PASS-WPT, but has no PASS gain. The transmit power is minimized by optimizing the transmit beamforming while guaranteeing the SINR requirement of the users. The optimization problem can be equivalently transformed into a convex problem and can be solved by ADMM. For simplicity, the conventional MIMO with SWIPT is represented by MIMO. 
    \item \textbf{PASS Enabled SWIPT}\cite{passSWIPT}: The baseline PASS follows the existing PASS design without power radiation model. The power is equally allocated to all PAs along the waveguide, i.e., $\alpha_{n,l} = \sqrt{1/L}$. The WMMSE algorithm is adopted by this baseline to solve the optimization problem. It is represented by PASS to simply the legend.
    \item \textbf{Proposed PASS-WPT}: Joint optimization of beamforming $\mathbf{W}$, PA positions $\mathbf{X}$, and power radiation ratios $\bm{\alpha}$ at each discrete \(\mathbf{X}\) via the AO and WMMSE for two-user case and QT-LDT for multi-user case.
\end{itemize}

\begin{figure}[htbp]
    \centering
    \includegraphics[width=3.3in]{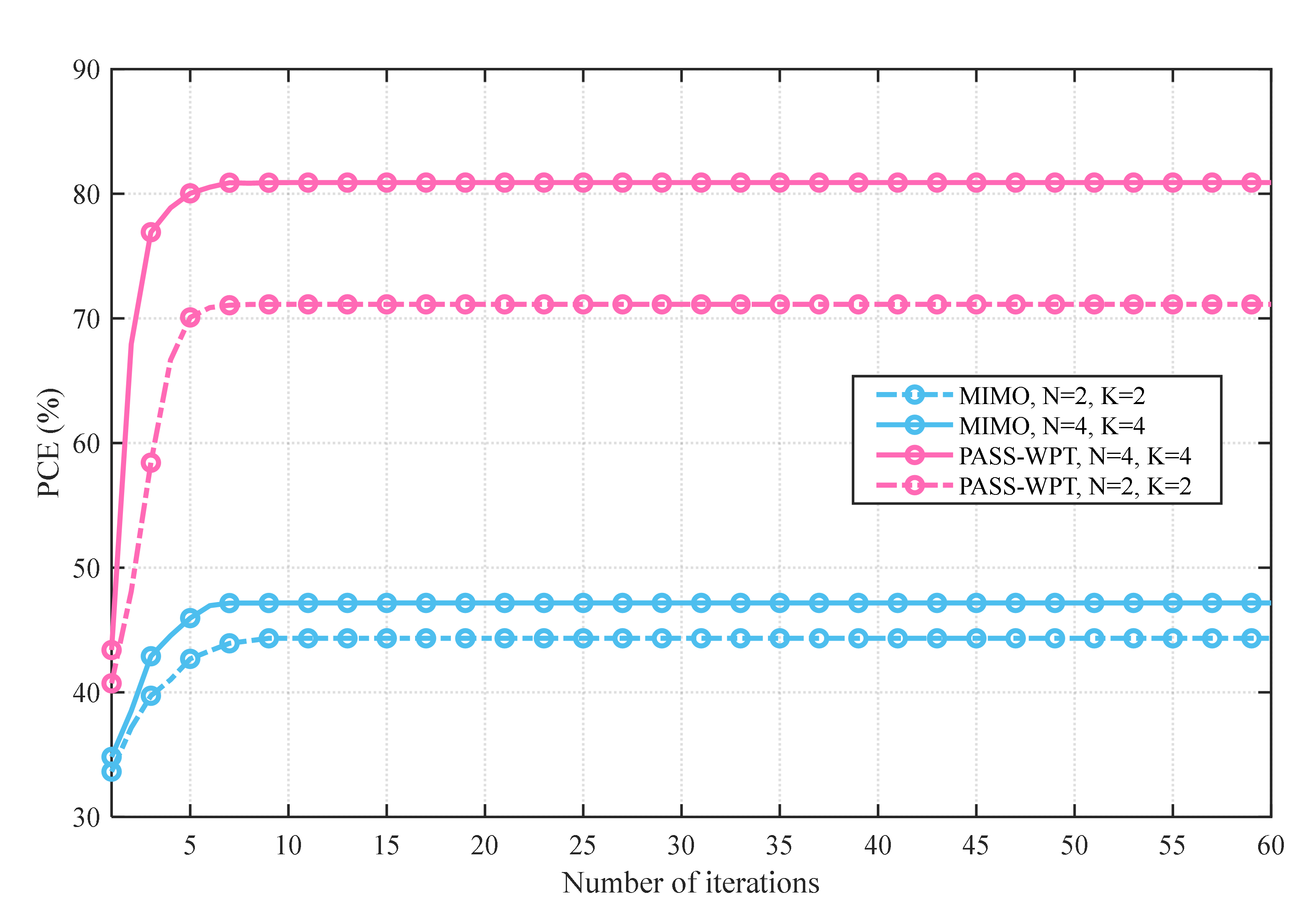}
    \caption{Comparisons of convergence behaviors of the proposed PASS-WPT and conventional MIMO.}
    \label{fig_2}
\end{figure}

Fig. \ref{fig_2} illustrates the convergence behaviors of PCE versus iterations for the proposed PASS-WPT and the conventional MIMO with $K=2, N=2$ and $K=4, N=4$. In all situations, PCE exhibits a rapid, monotone ascent and then quickly saturates. The proposed PASS-WPT converges within about 5 iterations, while conventional MIMO typically stabilizes after 8 iterations. The steady PCE value achieved 71.60$\%$ and 80.98$\%$ by PASS-WPT are markedly higher than the conventional MIMO with 43.71$\%$ and 45.73$\%$ improvements for 2 IDRs and 4 IDRs, respectively. These gains stem from the reconfigurable PASS design and optimized nonnegative power radiation ratios, which enhance energy focusing at EHRs increasing PCE.
Increasing $(N,K)$ from $(2,2)$ to $(4,4)$ raises the achievable PCE for both architectures, the proposed PASS-WPT is benefited more prominently. Because more degrees of freedom enable strong energy conversion and interference management.

\begin{figure}[htbp]
    \centering
    \includegraphics[width=3.3in]{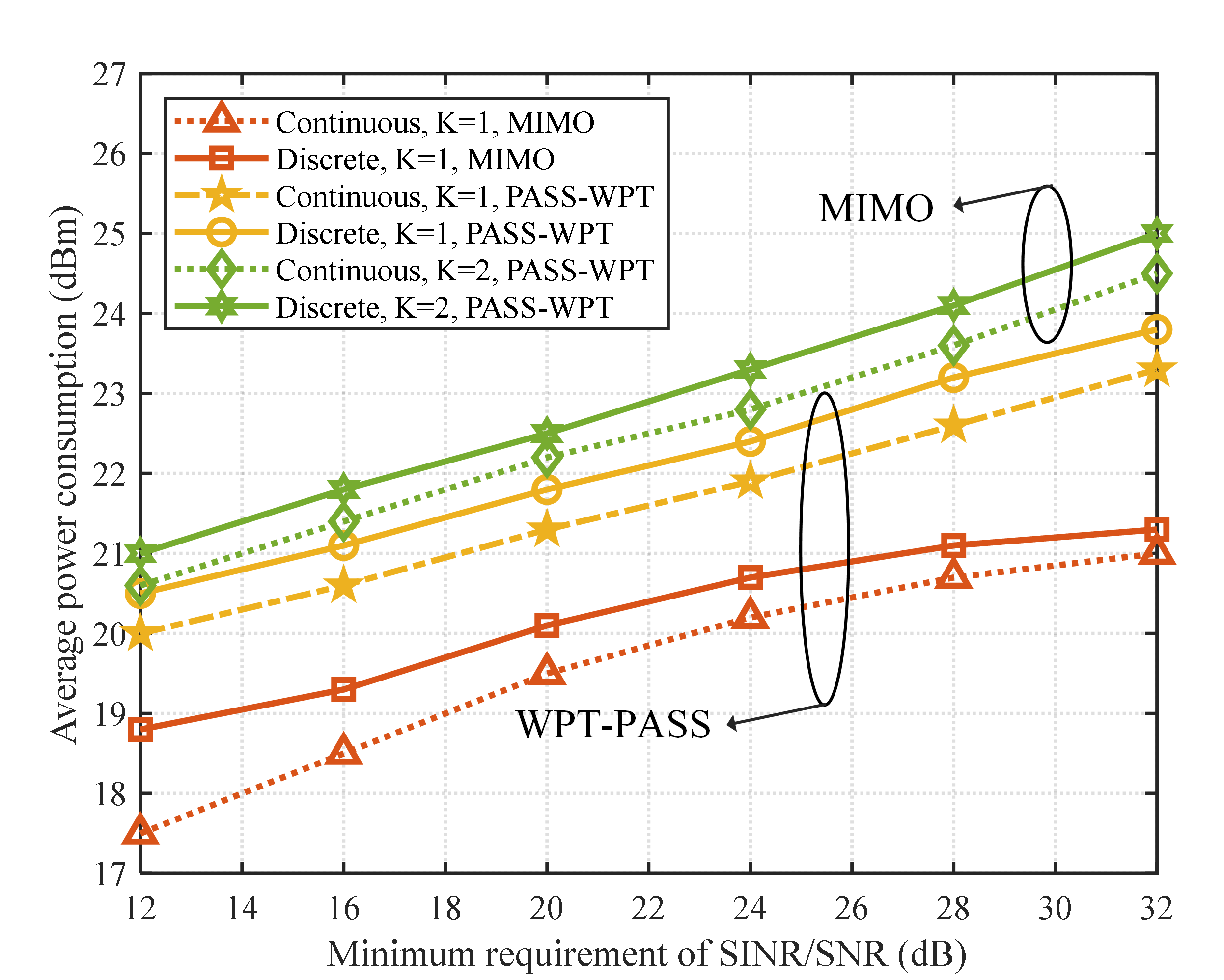}
    \caption{Comparisons of average power consumption of the proposed PASS-WPT and conventional MIMO.}
    \label{fig_3}
\end{figure}

Fig. \ref{fig_3} demonstrates average power consumption versus the minimum SINR/SNR requirement for continuous and discrete PA positions under two-user $(K=1, Q=1)$ and multi-user cases $(K=2, Q=2)$. Across all settings, as the minimum requirement of SINR/SNR, the average power consumption increases, approximately linearly with the target. Most notably, the proposed PASS-WPT outperforms conventional MIMO, and the gap widens as the SINR/SNR requirement grows. At 32 dB minimum requirement of SNR, the PASS-WPT with discrete positions decreases the average power consumption of the conventional MIMO with discrete positions by 16.82$\%$. The continuous implementation consistently outperforms the discrete one, owing to its finer alignment of the effective aperture and hence improving energy conversion. Multi-user case, e.g., $(K=2,Q=2)$, requires higher power than two-user case and exhibits a steeper slope, which is attributable to inter-user interference and the need to satisfy multiple user thresholds concurrently.

\begin{figure}[htbp]
    \centering
    \includegraphics[width=3.3in]{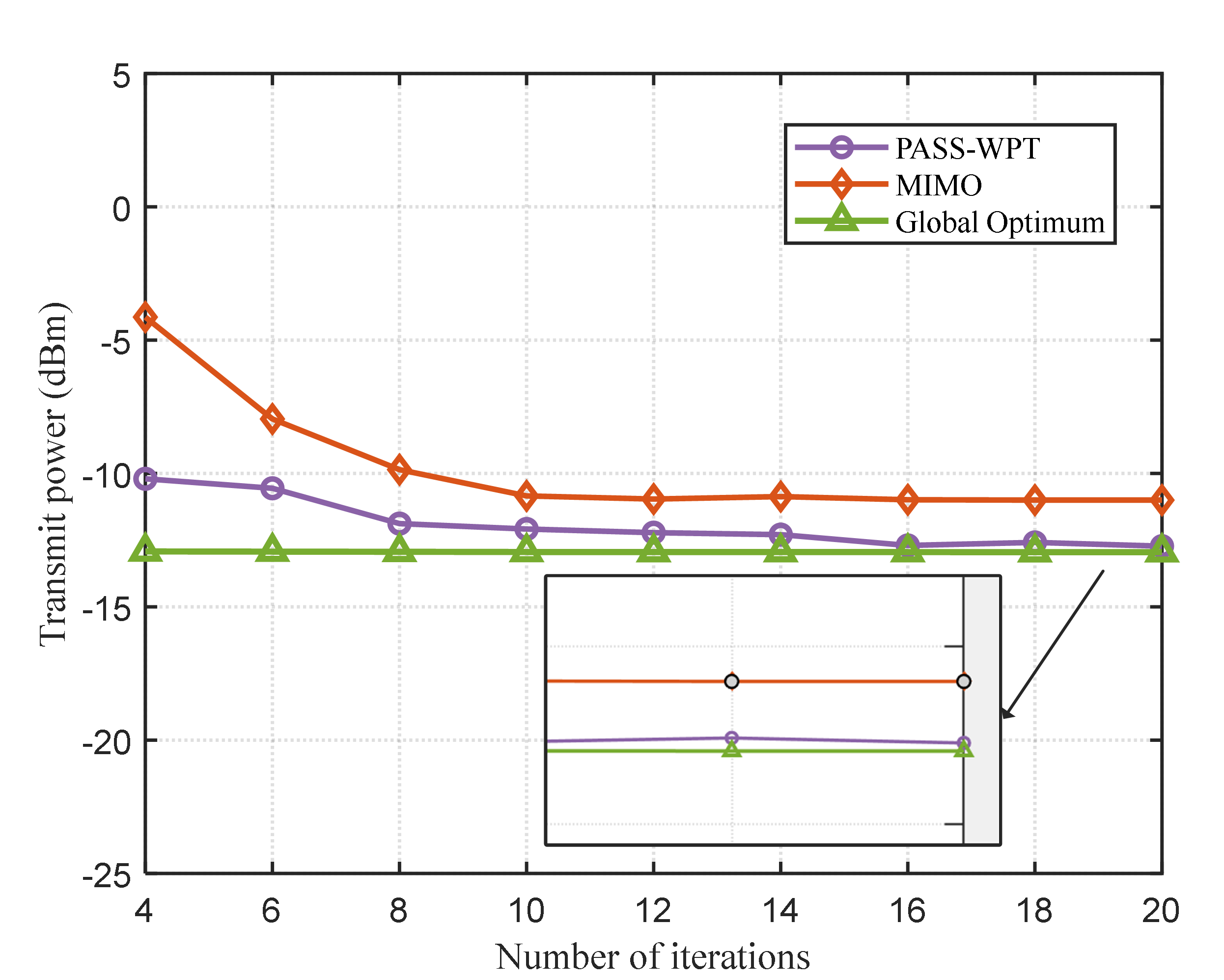}
    \caption{Comparisons of transmit power of the proposed PASS-WPT and conventional MIMO.}
    \label{fig_4}
\end{figure}

Fig. \ref{fig_4} depicts the transmit power of the proposed PASS-WPT and conventional MIMO versus the global optimum. When it iterates, both the proposed PASS-WPT and conventional MIMO decrease monotonically, and PASS-WPT converges faster and closer to the global optimum value. Quantitatively, at the 4-$th$ iteration, conventional MIMO requires about -4.5 dBm whereas PASS-WPT is near -10 dBm, which is about 5.5 dBm saving. When it converges at 20-$th$ iterations, the PASS-WPT is close to the global optimum within 0.35$\%$ errors while conventional MIMO has the 7.98$\%$ errors. Moreover, at convergence, PASS-WPT achieves about 2dB reduction over conventional MIMO which reduces by about 37.21$\%$ transmit power while remaining virtually optimal with respect to the global benchmark. These results substantiate that the proposed WPT design in PASS both accelerates convergence and delivers near-optimal power efficiency compared with conventional MIMO.

\begin{figure}[htbp]
    \centering
    \includegraphics[width=3.3in]{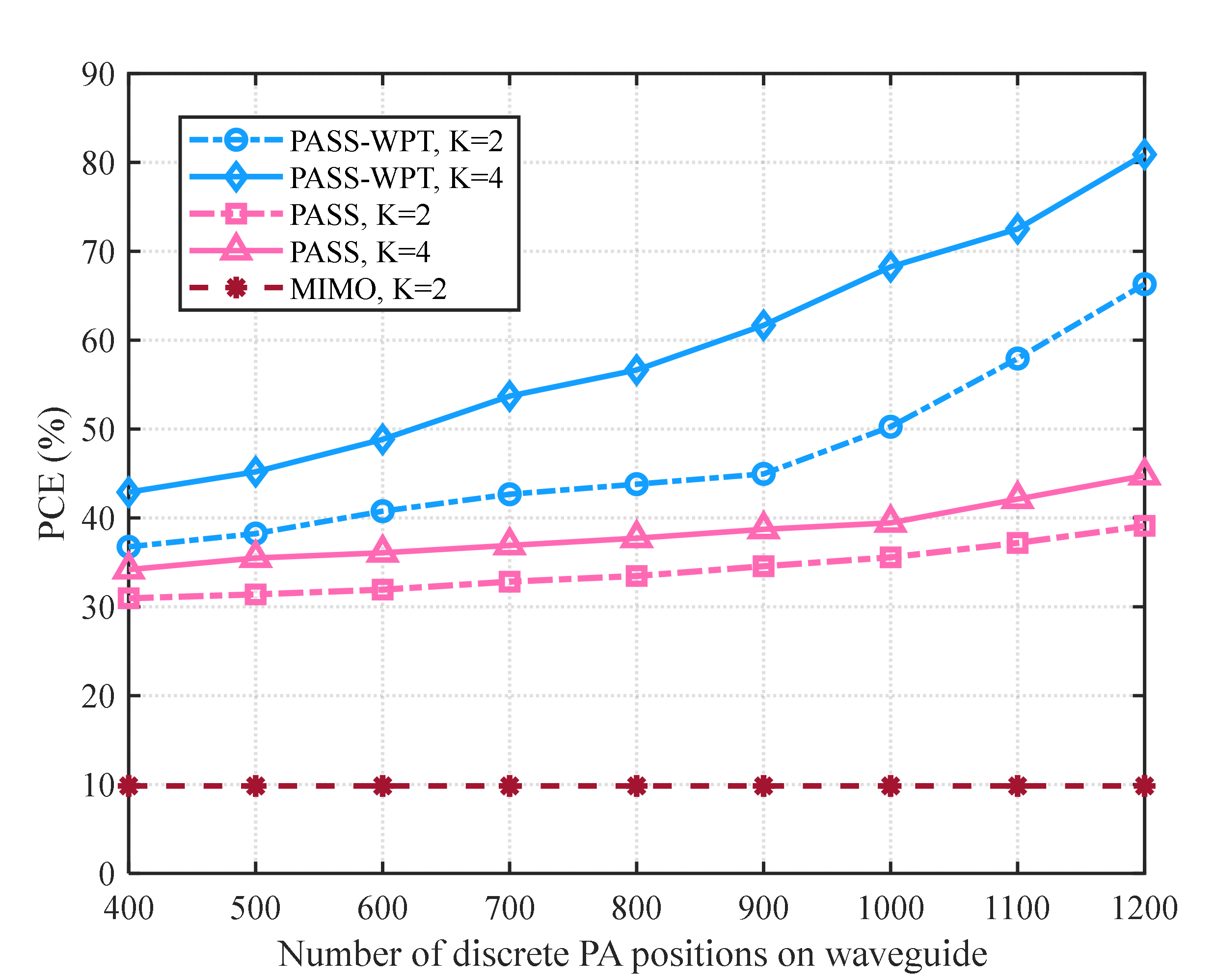}
    \caption{Comparisons of PCE of three frameworks versus discrete PA positions.}
    \label{fig_5}
\end{figure}

Fig. \ref{fig_5} illustrates the impact of the density of available discrete PA positions on each waveguide. As the grid becomes denser, the PCE increases monotonically because more PA positions enable fine-grained pinching beamforming optimization and more energy conversion at EHRs. In particular, enlarging the grid from 400 to 1200 positions raises the PCE of the proposed PASS-WPT framework from 37.32$\%$ to 66.38$\%$ for $(K=2,Q=2)$ and from 43.19$\%$ to 81.45$\%$ for $(K=4,Q=4)$, i.e., gains of 29.06$\%$ and 38.26$\%$, respectively. We note that such a large grid size substantially increases computational burden, since the outer selection over $\mathbf{X}$ scales linearly with the number of PA positions and each discrete position requires solving the inner convex subproblems. In summary, while the discrete implementation eases hardware deployment, it incurs a performance loss relative to the continuous ideal, PASS-WPT markedly mitigates this gap and benefits the most from increased grid density, especially for multi-user case.

\begin{figure}[htbp]
    \centering
    \includegraphics[width=3.3in]{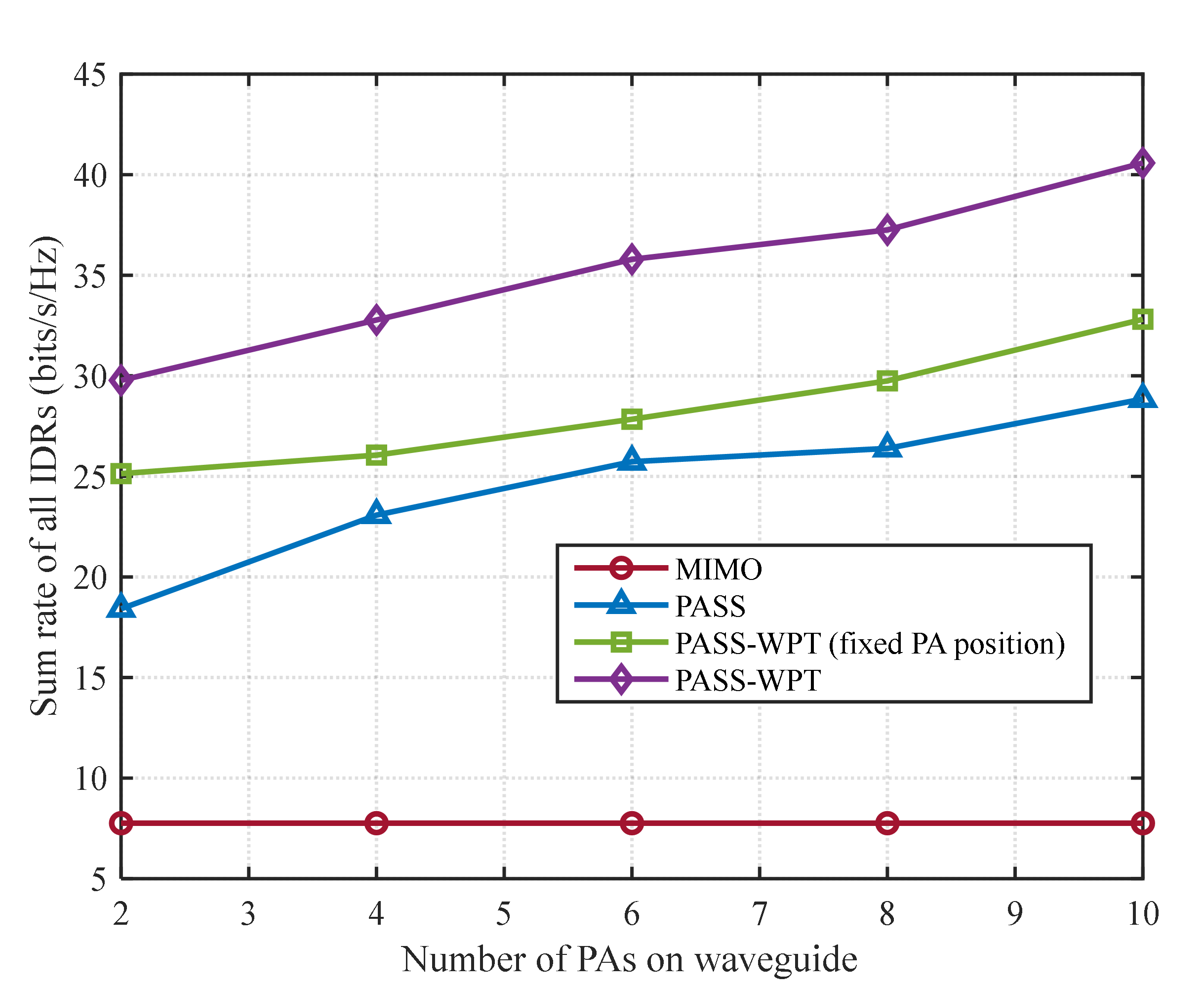}
    \caption{Comparisons of sum rate of IDRs versus number of PAs on waveguide.}
    \label{fig_6}
\end{figure}

Fig. \ref{fig_6} describes the sum rate of IDRs the number of PAs on each waveguide. 
As the PA increases, the sum rates of IDRs of all PASS frameworks increase. With 4 PAs on each waveguide, PASS-WPT achieves 77.81$\%$, 31.92$\%$, 23.08$\%$ improvements of sum rates compared to conventional MIMO, PASS, PASS-WPT with fixed positions. The widening gap from 2 to 10 PAs, indicates that optimizing PA locations and power radiation ratios, yields incremental benefits as the joint beamforming and power radiation optimization design in PASS-WPT provides the highest and most scalable throughput and energy conversion.

\begin{figure}[htbp]
    \centering
    \includegraphics[width=3.3in]{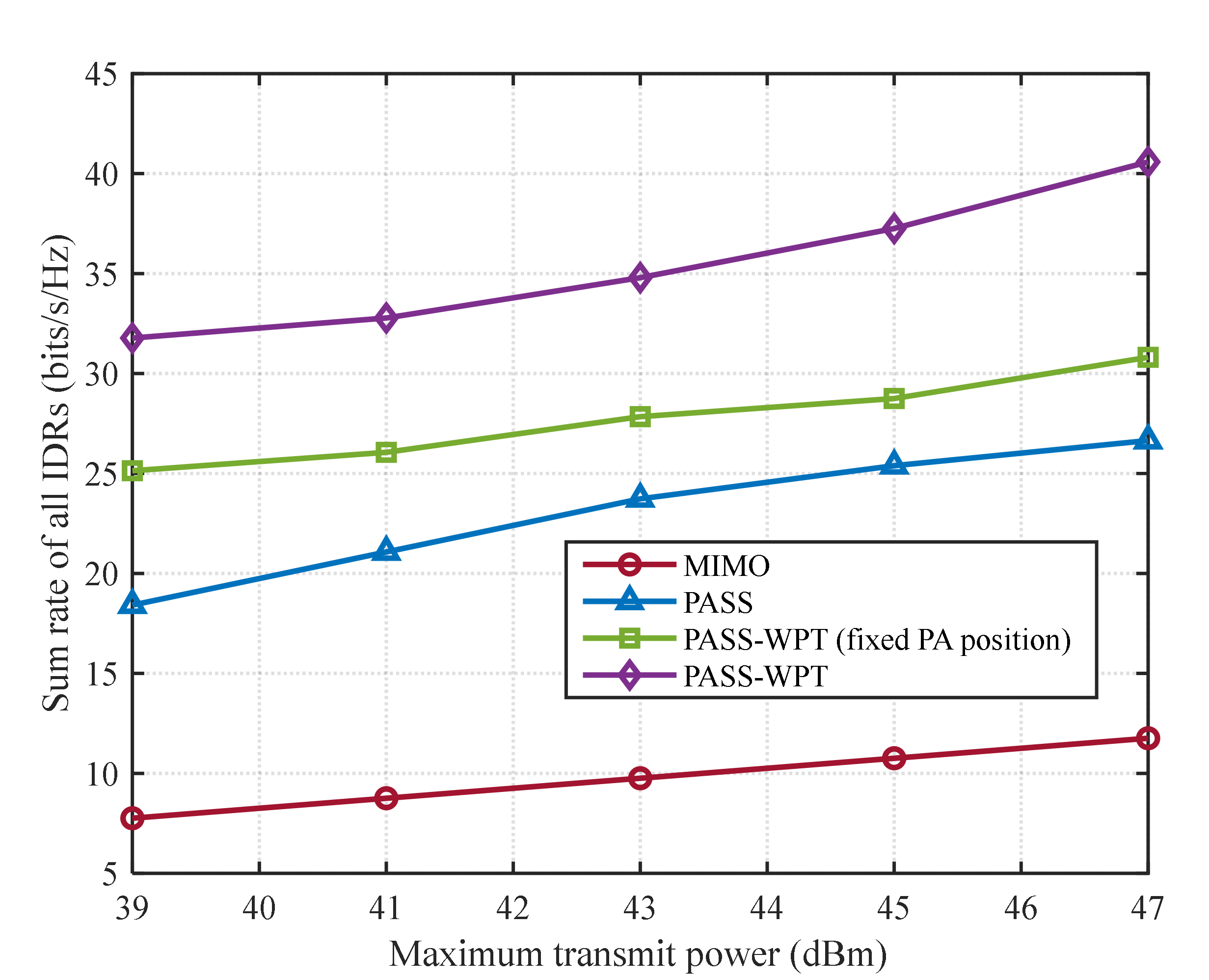}
    \caption{Comparisons of sum rate versus different transmit power.}
    \label{fig_7}
\end{figure}

Fig. \ref{fig_7} shows the sum rate of all IDRs versus the maximum transmit power for four schemes. All curves increase with maximum power budget, the sum rates of IDRs of four schemes increase. At $P^{\max}=$39 dBm, PASS-WPT improves the sum rate by 27.21$\%$, 69.1$\%$, 97.5$\%$ compared to PASS-WPT with fixed position, PASS, conventional MIMO. Since PASS-WPT not only attain the highest absolute throughput but also scales more favorably with the maximum transmit power budget, and optimizing power radiation ratios and PA positions provides an additional 31.00$\%$ gain.

\section{Conclusion}

In this paper, a novel PASS-WPT framework has been proposed for downlink MIMO scenarios, which integrates WPT with adjustable pinching beamforming and power radiation to enhance energy harvesting and communication performance.
To address the strong coupling among the above objectives, we have formulated a bi-level optimization problem, consists of the upper level PCE maximization problem and the lower level sum rate maximization problem. Specifically, the upper level PCE maximization problem has been proposed to optimize the transmit beamforming, PA position, and feasible interval of power radiation ratio. Then, the lower level sum rate maximization problem has been formulated to finetune the power radiation radio with the solutions of the upper level problem. We have developed closed-form solutions of the bi-level problems in two-user and multi-user cases.
For the two-user case, AO-based and WMMSE-based algorithms have been designed to derive closed-form solutions of transmit beamforming, PA positions, and power radiation ratios. For the multi-user case, a unified QT-LDT-based algorithm has been proposed to maximize upper level problem and lower level problem, which both have distributed the optimization process into two steps, where PA position update step for optimization of transmit beamforming, PA positions and feasible interval of power radiation ratio, and power radiation ratio update step for refining this ratio.
Extensive simulation results have proved the superiority of the proposed PASS-WPT framework over the conventional MIMO and classic PASS. Specifically, the proposed framework can achieve 81.45$\%$ and 43.19$\%$ improvements in PCE compared to conventional MIMO and baseline PASS, respectively. Moreover, the proposed framework can increase the sum rate by 77.81$\%$ and 31.91$\%$ relative to the conventional MIMO and PASS benchmarks. These simulation results have demonstrated that PASS-WPT provides a promising solution for the co-design of wireless information and power transfer in next-generation wireless networks.



\bibliographystyle{IEEEtran}
\def\baselinestretch{1}
\bibliography{WPT_PASS}

\end{document}